%% file: decoupling.tex
\PassOptionsToPackage{unicode}{hyperref}
\PassOptionsToPackage{hyphens}{url}
\PassOptionsToPackage{dvipsnames,svgnames*,x11names*}{xcolor}
\documentclass[
  astrosymb, twocolumn, tighten, twocolappendix]{aastex63}
\usepackage{amsmath,amssymb}
\usepackage{lmodern}
\usepackage{ifxetex,ifluatex}
\ifnum 0\ifxetex 1\fi\ifluatex 1\fi=0 
  \usepackage[T1]{fontenc}
  \usepackage[utf8]{inputenc}
  \usepackage{textcomp} 
\else 
  \usepackage{unicode-math}
  \defaultfontfeatures{Scale=MatchLowercase}
  \defaultfontfeatures[\rmfamily]{Ligatures=TeX,Scale=1}
\fi
\IfFileExists{upquote.sty}{\usepackage{upquote}}{}
\IfFileExists{microtype.sty}{
  \usepackage[]{microtype}
  \UseMicrotypeSet[protrusion]{basicmath} 
}{}
\makeatletter
\@ifundefined{KOMAClassName}{
  \IfFileExists{parskip.sty}{%
    \usepackage{parskip}
  }{
    \setlength{\parindent}{0pt}
    \setlength{\parskip}{6pt plus 2pt minus 1pt}}
}{
  \KOMAoptions{parskip=half}}
\makeatother
\usepackage{xcolor}
\IfFileExists{xurl.sty}{\usepackage{xurl}}{} 
\IfFileExists{bookmark.sty}{\usepackage{bookmark}}{\usepackage{hyperref}}
\hypersetup{
  colorlinks=true,
  linkcolor=Maroon,
  filecolor=Maroon,
  citecolor=Blue,
  urlcolor=Blue,
  pdfcreator={LaTeX via pandoc}}
\usepackage{graphicx}
\makeatletter
\def\maxwidth{\ifdim\Gin@nat@width>\linewidth\linewidth\else\Gin@nat@width\fi}
\def\maxheight{\ifdim\Gin@nat@height>\textheight\textheight\else\Gin@nat@height\fi}
\makeatother
\setkeys{Gin}{width=\maxwidth,height=\maxheight,keepaspectratio}
\makeatletter
\def\fps@figure{htbp}
\makeatother
\providecommand{\tightlist}{%
  \setlength{\itemsep}{0pt}\setlength{\parskip}{0pt}}
\input{macros.tex}
\usepackage{mathptmx,txfonts,tikz}
\ifluatex
  \usepackage{selnolig}  
\fi
\usepackage[]{natbib}
\begin{document}

\shorttitle{Mixed Modes and Surface Effects I}
\title{Mixed Modes and Asteroseismic Surface Effects: I. Analytic Treatment}
\input{preamble}
\begin{abstract}
Normal-mode oscillation frequencies computed from stellar models differ from those which would be measured from stars with identical interior structures, because of modelling errors in the near-surface layers. These frequency differences are referred to as the asteroseismic "surface term". The vast majority of solar-like oscillators which have been observed, and which are expected to be observed in the near future, are evolved stars which exhibit mixed modes. For these evolved stars, the inference of stellar properties from these mode frequencies has been shown to depend on how this surface term is corrected for. We show that existing parametrisations of the surface term account for mode mixing only to first order in perturbation theory, if at all, and therefore may not be adequate for evolved stars. Moreover, existing nonparametric treatments of the surface term do not account for mode mixing. We derive both a first-order construction, and a more general approach, for one particular class of nonparametric methods. We illustrate the limits of first-order approximations from both analytic considerations and using numerical injection-recovery tests on stellar models. First-order corrections for the surface term are strictly only applicable where the size of the surface term is much smaller than both the coupling strength between the mixed p- and g-modes, as well as the local g-mode spacing. Our more general matrix construction may be applied to evolved stars, where perturbation theory cannot be relied upon.
\end{abstract}

\defcitealias{ball_correction_2014}{BG14}
\defcitealias{grevesse_standard_1998}{GS98}
\defcitealias{ong_semianalytic_2020}{OB20}
\newcommand\bg{{\citetalias{ball_correction_2014}}}
\newcommand\gs{{\citetalias{grevesse_standard_1998}}}
\newcommand\ob{{\citetalias{ong_semianalytic_2020}}}
\newcommand\mesa{{\textsc{mesa}}}
\newcommand\gyre{{\textsc{gyre}}}

\hypertarget{introduction-and-motivation}{%
\section{Introduction and
Motivation}\label{introduction-and-motivation}}

High-cadence stellar photometry from space missions like \emph{CoRoT},
\emph{Kepler}, and \emph{TESS} (and soon CHEOPS, PLATO, and others) has
enabled the detection of solar-like oscillations in stars spanning a
broad range of evolutionary stages and masses, as well as the
measurement of their oscillation frequencies with high precision.
However, efforts to use these measurements to constrain the properties
of these stars, as was done to great effect for the Sun with
helioseismology, are fundamentally limited by deficiencies in modelling
their surface layers. As a result of these modelling errors, the
normal-mode oscillation frequencies computed from stellar models
necessarily differ from those which would be measured from stars with
identical global properties and interior structures. These frequency
differences are collectively referred to as the asteroseismic ``surface
term''. For p-modes, the surface term is understood to be a slowly
varying function of frequency, whose magnitude also increases with
frequency.

\citet{ong_differential_2021} showed that the behaviour of the surface
term exhibits qualitative differences between main-sequence stars and
more evolved (red giant) stars. In particular, they confirmed
\citep[following][]{basu_robustness_2018} that parametric methods like
the proposed correction of \citet[][hereafter \bg]{ball_correction_2014}
yield estimates of the stellar mass, radius, and age which are
comparable to those returned by nonparametric methods, such as
separation ratios
\citep{roxburgh_ratio_2003, otifloranes_use_2005, roxburgh_ratio_2005}
and phase offsets \citep{roxburgh_asteroseismic_2016}, when applied to
stars on the main sequence. However, they also demonstrated that the use
of these two nonparametric methods yielded ensemble estimates of these
same properties that differed from those obtained using the
\bg~correction, when applied to red giants in the open cluster NGC~6791.
Since many other parametric corrections --- motivated either by solar
observations \citep{kjeldsen_correcting_2008}, or from MHD simulations
\citep{sonoi_surface_2015} --- all yield estimates of these properties
which are similar to the \bg~correction on this open cluster
\citep{jorgensen_investigating_2020}, this was suggestive of a
qualitative difference between parametric and nonparametric treatments
of the surface term in these evolved oscillators. Furthermore, this
leaves open the possibility of some hitherto unexplored transition
between main-sequence-like and red-giant-like surface terms, as
reflected by how the results of parametric and nonparametric treatments
differ.

Ideally, we should investigate such a transition by examining potential
differences between both classes of surface-term treatments when applied
to stars in intermediate evolutionary phases (subgiants), since this
might be instructive as to when, or how, this qualitative change occurs
(or if there is a sharp transition at all). However, the oscillation
frequencies of subgiant stars exhibit significant, qualitative
differences from those of both main-sequence stars and more evolved red
giants. In particular, evolved solar-like oscillators support both
acoustic \(p\)-waves (in the convective exterior), and buoyancy
\(g\)-waves (in the radiative interior). Mathematically, these are
described by two independent sets of \(\pi\) and \(\gamma\) modes
\citep[in the sense of][]{aizenman_avoided_1977, ong_semianalytic_2020},
whose mode cavities are coupled to each other evanescently. The
normal-mode frequencies which we measure are not these ``bare'' \(\pi\)
and \(\gamma\) mode frequencies, but have been ``screened,'' both by
self-interaction (to yield \(p\) and \(g\)-modes), and by coupling to
each other to yield mixed modes.

In the two regimes which we have previously examined, we have
observational access to either pure \(p\)-modes (for main sequence
stars) or close to pure \(\pi\)-modes (for red giants). On the other
hand, subgiant stars, which lie between these two regimes, exhibit
isolated avoided crossings with very strong coupling between the \(\pi\)
and \(\gamma\) mode cavities, so that the quantities which are used in
nonparametric treatments of the surface term cannot easily be estimated
from their measured frequency sets. Consequently, these methods cannot
be directly applied to such subgiants. Likewise, the extent to which
existing parametrisations of the surface term, including that of \bg,
remain valid in the presence of mode mixing is still largely unexplored.

In this paper, we provide constructions generalising those of \bg~and of
\citet{roxburgh_asteroseismic_2016}, exploiting recent theoretical
developments \citep[][hereafter \ob]{ong_semianalytic_2020} permitting
the evaluation of the bare \(\pi\)- and \(\gamma\)-mode eigensystem of a
stellar model, as well as of their corresponding coupling matrices in
closed form. Using this, we examine the limits of validity of existing
approaches to the surface term in the presence of mode mixing
(\autoref{perturbation-analysis}). We also extend one class of
nonparametric treatments of the surface term \citep[the
\(\epsilon\)-matching algorithm of][]{roxburgh_asteroseismic_2016} to
explicitly take mode coupling into account, and examine various sources
of systematic error committed in our construction
(\autoref{epsilon-matching}). We then demonstrate the utility of these
generalisations in modelling stars exhibiting such isolated avoided
crossings, by employing these procedures in an injection-recovery test
on stellar models, and assess the practical significance of this
systematic error (\autoref{tests-using-stellar-models}). In the
companion paper to this work \citep[hereafter Paper II]{ong_surface_2},
we use the procedures outlined here to investigate the behaviour of the
surface term, characterised through these constructions, as seen in a
larger sample of various subgiants observed by the \emph{Kepler} and K2
missions.

\hypertarget{perturbation-analysis}{%
\section{Perturbation analysis}\label{perturbation-analysis}}

The avoided crossings with which we are concerned arise owing to the
coupling of two otherwise disjoint acoustic systems, whose frequencies
are close to resonance. In general, these frequencies are the
eigenvalues of some time-independent wave operator \(\hat{\mathcal{L}}\)
associated with the stellar structure. \ob~described a decomposition of
this wave operator into separate \(\pi\) and \(\gamma\) wave operators
\(\hat{\mathcal{L}}_\pi, \hat{\mathcal{L}}_\gamma\) --- constructed so
as to suppress wave propagation of the each type separately --- and
their remainder operators
\(\hat{\mathcal{R}}_\pi, \hat{\mathcal{R}}_\gamma\). Having done so,
they then derived analytic expressions for the matrix elements of a
generalised Hermitian eigenvalue problem (GHEP) of the form
\begin{equation}
  \begin{aligned}
  \begin{bmatrix}
  \mathbf{L}_\pi & \mathbf{L}_{\pi\gamma}\\
  \mathbf{L}_{\pi\gamma}^\dagger & \mathbf{L}_\gamma
  \end{bmatrix} \mathbf{c}_i 
  &=
  \begin{bmatrix}
  -\mathbf{\Omega}^2_\pi + \mathbf{R}_{\pi\pi} & -\mathbf{\Omega}^2_\pi\mathbf{D}_{\pi\gamma} + \mathbf{R}_{\pi\gamma}\\
  \left(-\mathbf{\Omega}^2_\pi\mathbf{D}_{\pi\gamma} + \mathbf{R}_{\pi\gamma}\right)^\dagger & -\mathbf{\Omega}^2_\gamma + \mathbf{R}_{\gamma\gamma}
  \end{bmatrix} \mathbf{c}_i 
  \\&=
   -\omega_i^2 \begin{bmatrix}
  \mathbb{I}_\pi & \mathbf{D}_{\pi\gamma}\\
  \mathbf{D}_{\pi\gamma}^\dagger & \mathbb{I}_\gamma
  \end{bmatrix}\mathbf{c}_i,
  \end{aligned}\label{eq:HEP}
\end{equation} yielding mixed-mode angular frequencies \(\omega_i\) and
mixing coefficients \(c_{ij}\) as the resulting eigenvalues and
eigenvectors, such that the mixed mode eigenfunctions are expressed as
linear combinations of those of \(\pi\) and \(\gamma\) modes with these
coefficients, as \(\xi_{\text{mixed},i} = \sum_j c_{ij}\xi_j\).
\amend{As in \ob, here $\mathbf{\Omega}_\pi$ and $\mathbf{\Omega}_\gamma$ are diagonal matrices whose entries are the squared angular frequencies of the $\pi$ and $\gamma$ modes, and $\mathbb{I}_\pi$ and $\mathbb{I}_\gamma$ are identity matrices of the same ranks as the number of $\pi$ and $\gamma$ modes under consideration.}

We adopt most of the expressions from \ob~for these matrix elements: in
particular, the volume integrals specifying the overlap terms
\begin{equation}
  {D_{\pi\gamma}}_{,ij} = \int \rho \ \vec{\xi}_{\pi,i}^* \cdot \vec{\xi}_{\gamma,j}\ \mathrm d^3 x,\label{eq:overlap}
\end{equation} where \(\vec{\xi}\) is the Lagrangian displacement
eigenfunction associated with a mode and \(\rho\) is the local density.
Likewise, we use their expressions for the \(\pi\)-mode interaction
terms: \begin{equation}
  {R_{\pi\pi}}_{,ij} = \left<\xi_{\pi,i}, \hat{\mathcal{R}}_\pi\vec{\xi}_{\pi, j}\right> = -\int \rho N^2\xi_{r, \pi,i}^*\xi_{r,\pi,j} ~ \mathrm d^3 x,\label{eq:Rpi}
\end{equation} where \(N^2\) is the squared Brunt-Väisälä frequency.
These expressions assume unit normalisation of the eigenfunctions with
respect to the standard inner product: \begin{equation}
    \int \rho\ \vec{\xi}_{\pi,i}^* \cdot \vec{\xi}_{\pi,j}~\mathrm d^3 x = \int \rho \ \vec{\xi}_{\gamma,i}^* \cdot \vec{\xi}_{\gamma,j}~\mathrm d^3 x = \delta_{ij}.
\end{equation} For the \(\gamma\)-mode self-interaction terms
\(\mathbf{R}_{\gamma\gamma}\), we have derived new expressions that are
manifestly Hermitian; this addresses a significant shortcoming in the
\ob~construction. While essential to our subsequent numerical
calculations, the derivation of these new expressions is not the central
focus of this work, and we leave the details of it to
\autoref{sec:gammagamma}.

\hypertarget{perturbation-theory-for-the-generalised-hermitian-eigenvalue-problem}{%
\subsection{Perturbation Theory for the Generalised Hermitian Eigenvalue
Problem}\label{perturbation-theory-for-the-generalised-hermitian-eigenvalue-problem}}

\label{sec:perturb}

Given a perturbed Hermitian eigenvalue problem of the form
\begin{equation}
  (\mathbf{H}_0 + \lambda \mathbf{V})\mathbf{c}_n = \varepsilon_n \mathbf{c}_n,\label{eq:SE}
\end{equation} Rayleigh-Schrödinger perturbation theory permits a
description of the perturbed eigenvalues \(\varepsilon_n\) and
eigenvectors \(\mathbf{c}_n\) in terms of the eigensystem associated
with the unperturbed operator \(\mathbf{H}_0\), in the form of an
asymptotic series in powers of the parameter \(\lambda\). We refer the
reader to textbooks on linear algebra or quantum mechanics for reminders
of the usual expressions, which we will reproduce here without proof:
given \begin{equation}
  \varepsilon_n \sim \sum_k \lambda^k \varepsilon^{(k)}_n; (\mathbf{c}_n)_m \equiv c_{nm} \sim \sum_k \lambda^k c^{(k)}_{nm}, \label{eq:SE2}
\end{equation} the terms in this series expansion are found by expanding
\cref{eq:SE2} into \cref{eq:SE}, collecting terms by powers of
\(\lambda\), and demanding that the perturbed eigenvectors be
orthonormal. This results in a set of recurrence relations which yield
at last that \begin{equation}
\begin{aligned}
\varepsilon_n &\sim \varepsilon_n^{(0)} + \lambda V_{nn} + \lambda^2 \left(\sum_{m \ne n}{V_{nm}V_{mn} \over \varepsilon^{(0)}_n - \varepsilon^{(0)}_m}\right) + \mathcal{O}(\lambda^3),\\
c_{nm}^{(0)} &\sim \delta_{nm} + \lambda {V_{mn} \over \varepsilon^{(0)}_n - \varepsilon^{(0)}_m} + \mathcal{O}(\lambda^2).\label{eq:normalperturb}
\end{aligned}
\end{equation}

However, with respect to the bare \(\pi\) and \(\gamma\) modes,
\cref{eq:HEP} is a GHEP, where the right-hand-side of the equation
(proportional to the eigenvalues) is also potentially affected by the
perturbation: hence \cref{eq:normalperturb} is not directly applicable.
We regroup terms, and rewrite the eigenvalues as
\(-\omega_i^2 \equiv \varepsilon_i\), to yield \begin{equation}
\begin{aligned}
(\mathbf{L}_0 + \kappa \mathbf{P} + \lambda \mathbf{V})\mathbf{c}_n &=
 \begin{bmatrix}
  \mathbf{E}_\pi + \kappa\mathbf{P}_{\pi\pi} + \lambda \mathbf{V}_\text{surf} & \kappa\mathbf{P}_{\pi\gamma}\\
  \kappa\mathbf{P}_{\pi\gamma}^\dagger & \mathbf{E}_\gamma + \kappa\mathbf{P}_{\gamma\gamma}
  \end{bmatrix}
  \begin{bmatrix}
  \mathbf{c}_\pi \\ \mathbf{c}_\gamma
  \end{bmatrix}\\
  & = - \omega_n^2 \begin{bmatrix}
  \mathbb{I}_\pi + \lambda \mathbf{Q}_\text{surf} & \kappa\mathbf{D}_{\pi\gamma} \\ \kappa\mathbf{D}_{\pi\gamma}^\dagger & \mathbb{I}_\gamma
  \end{bmatrix} \begin{bmatrix}
  \mathbf{c}_\pi \\ \mathbf{c}_\gamma
  \end{bmatrix} \equiv \varepsilon_n \left(\mathbf{1} + \kappa\mathbf{D} + \lambda \mathbf{Q}\right)\mathbf{c}_n.
\end{aligned}\label{eq:perturb}
\end{equation} We have regrouped terms in order to treat \(\kappa\) and
\(\lambda\) as two separate parameters, both taking values between 0 and
1. The parameter \(\kappa\) quantifies the overall coupling between the
\(\pi\) and \(\gamma\) subsystems (described by the matrix
\(\mathbf{P}\)). As we reduce to the uncoupled problem with
\(\kappa \to 0\), the matrix \(\mathbf{L}_0\) on the LHS becomes
diagonal; its elements are given by the oscillation frequencies of the
bare \(\pi\) and \(\gamma\) modes. In the same limit, the ``metric''
matrix on the RHS reduces to the identity matrix. On the other hand,
\(\lambda\) describes the size of a structural perturbation to the
stellar model associated with some differential operator
\(\hat{\mathcal{V}}\); by convention, it also takes values between 0 and
1. For reasons that we will discuss in the following section, the
matrices \(\mathbf{V}\) and \(\mathbf{Q}\) can be assumed to vanish
outside the \(\pi\)-mode subspace when the perturbation represents the
surface term; however this property is not essential to our subsequent
discussion.

In principle, it is possible to choose some basis that diagonalises the
matrix \(\left(\mathbb{I} + \kappa \mathbf D\right)\) on the RHS for any
value of \(\kappa\), permitting the direct use of
\cref{eq:normalperturb}; this is the strategy which was pursued in \ob.
However, in this basis the diagonal elements of the LHS matrix
\(\mathbf{L}_0\) may lose their intuitive interpretations, as given
above. Moreover, since numerically we only have access to incomplete
matrices (as the underlying differential operators are of infinite
rank), in practice this change of basis can only be performed
approximately. This approximation must be performed to the same order of
accuracy as the order to which the perturbative expansion itself is
truncated. The resulting procedure becomes highly cumbersome very
quickly beyond first order.

To avoid this, we perform a Rayleigh-Schrödinger-like expansion to an
asymptotic series in \(\lambda\) in the usual fashion, under the
assumption of dominant balance that \(\kappa \sim \lambda\) (since
formally both take values between 0 and 1). Conceptually, we are
``turning on'' both the mixed-mode coupling and the structural
perturbation at the same time. This gives us the following explicit
expressions:

\begin{itemize}
\tightlist
\item
  At zeroth order in \(\lambda\) we require
  \(\varepsilon_n(\lambda = 0) = \varepsilon^{(0)}_n\) and
  \(c^{(0)}_{nm} = \delta_{mn}\), as in \cref{eq:normalperturb}.
\item
  At first order in \(\lambda\) we have \begin{equation}
   \varepsilon^{(1)}_n = V_{nn} + P_{nn}; \ c^{(1)}_{nm} = {V_{mn} + P_{mn} \over \epsilon^{(0)}_n - \epsilon^{(0)}_m} - D_{mn} - Q_{mn}\text{ for $m \ne n$}. \label{eq:firstorder}
  \end{equation} The self-mixing terms \(c^{(1)}_{nn}\) are found by
  demanding that
  \(\mathbf{c}_n^\dagger \left(\mathbf{1 + \lambda \left(D + Q\right)} \right)\mathbf{c}_n = 1 + \mathcal{O}(\lambda^2),\)
  yielding at last that \(c^{(1)}_{nn} = -D_{nn} - Q_{nn} = -Q_{nn}\).
\item
  Higher-order terms at \(k^\text{th}\) order in \(\lambda\) can be
  found recursively from those of lower order, through the recurrence
  relation \begin{equation}
  \begin{aligned}
  \left(\varepsilon^{(0)}_n - {\varepsilon^{(0)}_m}\right)\left(c^{(k)}_{nm} + \sum_l A_{ml}c^{(k-1)}_{nl}\right) + \varepsilon_n^{(k)}\delta_{mn} \\ = \sum_{l} B_{ml}c_{nl}^{(k-1)} - \sum_{j=1}^{k-1}\varepsilon_n^{(j)}\left(c_{nm}^{(k-j)} + \sum_l A_{ml}c_{nl}^{(k-j-1)}\right),\label{eq:recurrence1}
  \end{aligned}
  \end{equation} to produce the eigenvalue perturbation
  \(\varepsilon_n^{(k)}\) and eigenvector components \(c_{nm}^{(k)}\)
  where \(m \ne n\). Here we have written \(A_{mn} = D_{mn} + Q_{mn}\)
  and \(B_{mn} = V_{mn} + P_{mn}\) for brevity, since each pair of
  matrices appears together in \cref{eq:perturb}. The constraint of
  orthonormality gives \begin{equation}
  \begin{aligned}
  2 \mathrm{Re}\left(c_{nn}^{(k)}\right) + \sum_{l=1}^{k-1}\sum_m {c_{nm}^{(l)}}^*c_{nm}^{(k-l)} &+ 2\mathrm{Re}\left(\sum_m A_{nm} c_{nm}^{(k-1)}\right) \\&+ \sum_{l=1}^{k-2}\sum_{j,m} {c_{nj}^{(l)}}^*A_{jm} c_{nm}^{(k - l - 1)}= 0,\label{eq:recurrence2}
  \end{aligned}
  \end{equation} which then yields the self-mixing eigenvector
  components \(c_{nn}^{(k)}\).
\end{itemize}

It can be seen that these recurrence relations reduce to the usual
Rayleigh-Schrödinger expressions as \(A_{ij} \to 0\).

\hypertarget{the-surface-term-as-a-matrix-perturbation}{%
\subsection{The surface term as a matrix
perturbation}\label{the-surface-term-as-a-matrix-perturbation}}

\label{sec:bg14matrix}

The matrix construction of \ob~associates linear operators
\(\hat{\mathcal A}\) with matrix elements \(A_{ij}\) with respect to
some set of basis functions. These matrix elements are found by
performing integrals of the form \begin{equation}
  A_{ij} = \left<\vec\xi_i, \hat{\mathcal{A}} \vec\xi_j\right> = \int \mathrm d m\ \vec\xi_i^* \cdot \hat{\mathcal{A}} \vec\xi_j, \label{eq:matelements}
\end{equation} i.e.~using the inner product for which our chosen sets of
basis functions --- in this case, the isolated \(\pi\) and
\(\gamma\)-mode eigenfunctions --- are orthonormal. With respect to this
construction, we now consider frequency perturbations associated with
the surface term to result from a perturbation to the wave operator of
the form \begin{equation}
   \hat{\mathcal L} \to \hat{\mathcal L} + \lambda \hat{\mathcal V}.
\end{equation} The operator \(\hat{\mathcal{V}}\) carries the following
interpretation: we consider two different stellar structures, which we
constrain to have identical global properties (in particular, identical
mass and radius). We assume we have access to the eigensystem of one of
them, which is associated with the wave operator \(\hat{\mathcal{L}}\);
we treat this as our fiducial structure (or reference model, in the
parlance of helioseismology). The second stellar structure has a
different set of pulsation frequencies and eigenfunctions, associated
with a different wave operator, which we write as
\(\hat{\mathcal{L}} + \hat{\mathcal{V}}\), acting on the same domain.
The parameter \(\lambda\), taking values in the range \([0, 1]\), serves
to interpolate between the two structures. In this manner, we may
express differences between two stellar structures as a classical
operator (and therefore matrix) perturbation problem, parameterised by
\(\lambda\).

Accordingly, characterising the surface term within this matrix
construction requires computing the matrix elements of
\(\hat{\mathcal V}\) via \cref{eq:matelements}, with respect to the
basis set of eigenfunctions from the fiducial model. If
\(\hat{\mathcal{V}}\) represents the surface term --- i.e.~the
structural difference compared to our fiducial model is localised to the
stellar surface --- then by assumption, we should also have
\begin{equation}
  \hat{\mathcal V} \sim \delta(r - R) \implies \hat{\mathcal V} \xi_{\gamma,i} \to 0,\label{eq:gammasmall}
\end{equation} since \(\gamma\) modes are confined to the stellar
interior, and thus in principle are unaffected by the surface term. By
an analogous argument, it can be shown that the matrix elements
\(Q_{ij}\) also vanish when evaluated with respect to \(\gamma\)-modes.

Based on these properties, we now construct an analogous parametrisation
to \bg~within this matrix formalism. In particular, we observe that the
diagonal elements of \cref{eq:HEP} can be rewritten as integrals of the
form \begin{equation}
  \omega_i^2 \int \mathrm d m\ \vec\xi_i \cdot \vec\xi_i + \int \mathrm dm\ \vec{\xi}_i \cdot \hat{\mathcal{L}} \vec\xi_i = 0.\label{eq:variational}
\end{equation} Not coincidentally, this is precisely the same structure
as the ``variational'' construction employed by
\citet{lyndenbell_stability_1967} and subsequently by
\citet{gough_comments_1990}, neglecting the rotational splitting term,
from which \bg~derive their ansatz parametrisation. In recovering
frequency perturbations from such integrals, these approaches restrict
consideration to only the diagonal elements of the matrix construction.

Obversely, we know that the diagonal elements of the perturbation matrix
completely specify the perturbations to the frequency eigenvalues only
to first order in the Rayleigh-Schrödinger expansion;
cf.~\cref{eq:normalperturb}. The variational analysis presented by
\citet{gough_comments_1990} and elsewhere can therefore be interpreted
as the truncation of the perturbative expansion for the frequency
eigenvalues to leading order in the expansion parameter \(\lambda\).
Conversely, any generalisation to this first-order approach must require
that the off-diagonal matrix elements be specified. Retracing the
arguments of \citet{gough_comments_1990} (in particular, that
\(\nabla \cdot \vec{\xi}_i \sim \omega^2 \xi_{i,r}\) and
\(\xi_r \sim |\vec{\xi}|\) near the stellar surface), we find that
perturbations to the stellar model localised at the stellar surface
result in perturbation matrix elements of the form \begin{equation}
  V_{ij} \sim \left(\oint \mathrm d \Omega \ \vec{\xi}^*_i(R) \cdot \vec\xi_j(R) \right) \left(a + b\left( \omega_i^2 \omega_j^2 \over \omega_0^4\right)\right),
\end{equation} for some constants \(a, b, \omega_0\). This is a bilinear
form in the frequency eigenvalues, which reduces to the quadratic form
derived in \citet{gough_comments_1990} when evaluated along the diagonal
elements. The two coefficients \(a, b\) here correspond to the
parameters \(a_{-1}, a_3\) of \bg, which takes the form \begin{equation}
    \delta \nu_{nl} \sim \left.\nu_0 \left(a_{-1}\left(\nu_{nl} \over \nu_0\right)^{-1}+a_3\left(\nu_{nl} \over \nu_0\right)^3\right) \right/ I_{nl}.
\end{equation} In terms of those parameters, we may equivalently write
\begin{equation}
    V_{ij} \sim \left. -2\omega_0^2 \left(a_{-1} + a_{-3} \left(\omega_i^2 \omega_j^2 \over \omega_0^4 \right)\right) \right/ \sqrt{I_i I_j},
\end{equation} which is the appropriate matrix generalisation of the
correction of \bg. We note that \(\gamma\)-mode eigenfunctions are
evanescent outside of the central buoyant cavity, so this construction
also approximately satisfies \cref{eq:gammasmall}. In a similar manner,
we find that \begin{equation}
    Q_{ij} = \oint \mathrm d \Omega \ \delta\rho(R) \ \vec{\xi}^*_i(R) \cdot \vec\xi_j(R),
\end{equation} where \(\delta\rho\) is the static localised perturbation
to the density profile associated with the surface term. This term gives
rise to a contribution to the frequency perturbation that goes as
\(\delta\omega_{nl} \sim \omega_{nl} / I_{nl}\), which is usually
assumed to be negligible in analyses of the surface term. We will also
neglect it in our subsequent discussion.

\amend{
While we have limited ourselves to considering only structural perturbations, to preserve the analogy with \cite{gough_comments_1990,ball_correction_2014}, we note that this matrix construction is not strictly limited to them. For example, interactions with a magnetic field or a nonzero fluid velocity field may also result in additional force terms in the momentum equation. In such cases, the linearised contributions of these terms in the time-independent wave equation, \cref{eq:waveorig}, take the form of additional differential operators acting on the wavefunctions \citep[e.g.][]{lyndenbell_stability_1967,gough_rotation_1990}, perturbing the fiducial system in a similar manner. Their associated matrix elements can also be found by evaluating volume integrals per \cref{eq:matelements}, and the above analysis can then be applied wholesale (up to modified parameterisation), potentially also incorporating additional dependences on the azimuthal order $m$ through a quadratic variant of the GHEP (cf. Ong \& Basu, in prep.). However, the details of how these matrix corrections are parametrised will still depend on the physical assumptions being made, just as is the case with pure p-modes. It is precisely for this reason that a nonparametric surface-term diagnostic remains desirable.
}

\hypertarget{convergence-of-the-perturbative-expansion}{%
\subsection{Convergence of the perturbative
expansion}\label{convergence-of-the-perturbative-expansion}}

\label{sec:convergence}

A formal expansion in powers of \(\lambda\) yields good approximations
only if the perturbation is heuristically ``small'', since the resulting
series is asymptotic but not necessarily convergent. On the other hand,
as we have just discussed, using only diagonal matrix elements to
describe the surface term coincides with the truncation of this series
expansion to leading order in \(\lambda\). This is more generally also
true of any result obtained with ``variational'' methods. Therefore,
determining the conditions under which this perturbative expansion
converges will also illuminate the limits of validity for these existing
methods.

To proceed, we need to first establish some properties of the
perturbation coefficients \(\varepsilon^{(k)}_n\). When the coupling is
weak enough to allow us to neglect the overlap terms \(D_{ij}\), we may
recycle standard expressions that apply to the Rayleigh-Schrödinger
expansion \cref{eq:normalperturb}, with \(B\) in place of \(V\). By
inspection of \cref{eq:recurrence1}, it can be shown that for \(k > 2\),
these can be written as a sum in powers of \(B_{nn}\): \begin{equation}
\begin{aligned}
\varepsilon_n^{(k)} &\sim \left(-B_{nn}\right)^{k-2}\sum_{m \ne n} {{|B_{mn}|}^2 \over \left(\varepsilon^{(0)}_{n} - \varepsilon^{(0)}_{m}\right)^{k-1}} \\
&+ \left(-B_{nn}\right)^{k-3} \sum_{m_1,m_2 \ne n}\sum_{l=1}^{k-1}{B_{nm_1}B_{m_1m_2}B_{m_2n} \over \left(\varepsilon^{(0)}_{n} - \varepsilon^{(0)}_{m_1}\right)^l\left(\varepsilon^{(0)}_{n} - \varepsilon^{(0)}_{m_2}\right)^{k-l-1}}\\
&\vdots\\
&+ \sum_{m_1 \ne n}\cdots\sum_{m_{k-1} \ne n} {B_{nm_1} B_{m_1m_2}\cdots B_{m_{k-1} n} \over \left(\varepsilon^{(0)}_{n} - \varepsilon^{(0)}_{m_1}\right)\cdots\left(\varepsilon^{(0)}_{n} - \varepsilon^{(0)}_{m_{k-1}}\right)}.\label{eq:eigk}
\end{aligned}
\end{equation}

Let us first consider the case of a surface perturbation acting on
\(p\)-modes, or bare \(\pi\)-modes, where \cref{eq:normalperturb}
describes the perturbative expansion; correspondingly we use
\cref{eq:eigk} with \(V\) in place of \(B\). Since these are typically
observed at high radial order (\(n_p \gtrsim 10\)), the relative change
in the matrix elements \(V_{nm}\), keeping \(n\) fixed, is small as we
move off the diagonal, compared to the relative changes in the resonance
terms \(\varepsilon^{(0)}_{n} - \varepsilon^{(0)}_{m}\). Accordingly, we
approximate \(V_{nm} \sim V_{nn}\) near the diagonal. At the same time,
this sum is also dominated by terms near the diagonal, since
off-diagonal terms are otherwise heavily suppressed by the resonance
factors in the denominator. For any given \(p\) or \(\pi\)-mode, we have
\begin{equation}
  \left|\varepsilon_n^{(k)}\right| \lesssim k^q \max_{m\ne n}\left| {V_{nm}^k \over \left(\varepsilon^{(0)}_{n} - \varepsilon^{(0)}_{m}\right)^{k-1}}\right| \sim {k^q \left|V_{nn}\right|^k \over \left|\min_{m \ne n} \left(\varepsilon^{(0)}_{n} - \varepsilon^{(0)}_{m}\right)\right|^{k-1}},\label{eq:bound1}
\end{equation} where \(k^q\) counts the number of such terms that
survive the summation over alternating signs. If we neglect the
alternating signs, we have \(q\sim2\), which bounds the sum from above.
However, the smallest possible separation between \(p\)-modes or
\(\pi\)-modes in the asymptotic regime is roughly given by \(\Dnu\).
Writing
\(V_{nn} \sim \delta\omega_q^2 \sim 2\omega_n \cdot 2\pi\delta\nu_\text{surf}\),
we find \begin{equation}
  \left|\varepsilon_n^{(k)}\right| \lesssim k^2 \cdot |V_{nn}| \cdot \left(\delta\nu_\text{surf} \over \Dnu\right)^{k-1}.
\end{equation} In practice, the size of the frequency perturbation from
the surface term is small enough that it may also be expressed as a
phase shift in the eigenvalue equation (i.e.~a small fraction of
\(\Dnu\)). The ratio in the parentheses is then, by assumption, much
less than 1, and this series can be shown to converge via e.g.~the
integral test.

Turning our attention to mixed modes, we split our analysis into two
cases by comparing the matrix elements of \(\mathbf{V}\) and
\(\mathbf{P}\). Since the surface term leaves \(\gamma\) modes
unaffected, following our discussion in \autoref{sec:bg14matrix}, we
have \(V_{ij} \ll P_{ij}\) where \(i\) indexes a \(\pi\) mode and \(j\)
a \(\gamma\) mode. Where the coupling is strong, we also have
\(V_{i_1i_2} \ll P_{i_1j}\) for all modes \(i_n\) in the \(\pi\)-mode
subspace. Since we cannot neglect the mode coupling in these cases, we
cannot use \cref{eq:eigk} directly. Instead we must consider the coupled
system (including the avoided crossing) to be our fiducial set of
eigenvalues and eigenfunctions, so that the perturbative series is
described by \cref{eq:normalperturb}, and we again use \cref{eq:eigk}
with \(V\) in place of \(B\). A further complication is that the matrix
elements of \(V\) must now be evaluated with respect to the mixed-mode
rather than bare \(\pi\)-mode system. Once again we approximate
\(V_{i_1i_2} \sim V_{i_1i_1}\) within the \(\pi\)-mode subspace. For
\(\pi\)-dominated mixed modes, this leads us to also approximate
\(V_{mn} \sim V_{nn}\), whence we again obtain \cref{eq:bound1}.

In this case, however, the minimum distance between eigenvalues is not
solely determined by the asymptotic properties of the underlying
eigenvalues, since there are two sets of underlying asymptotic relations
which evolve independently of each other. Instead, as modes from the
\(\pi\) set come into resonance with modes from the \(\gamma\) set, the
minimum separation between modes in the resulting avoided crossing will
be determined by the coupling strength. This coupling strength is a
function of the mode frequency, which (assuming it varies slowly with
frequency) we approximate as some constant \(P\). Conversely, if a
resonant pair exists, then for each of them the sum over terms
\amend{in} \cref{eq:eigk} will be dominated by resonance terms involving
the other. Therefore, in these cases we obtain as our condition for
convergence that \begin{equation}
  \left|\varepsilon_n^{(k)}\right| \lesssim k^2 \cdot |V_{nn}| \cdot \left(\delta\nu_\text{surf} \over P / 8\pi^2\nu_\pi\right)^{k-1} \implies {\delta\nu_\text{surf} \over P / 8\pi^2\nu_\pi} \ll 1.\label{eq:bound2}
\end{equation} But this is essentially a restatement of our original
condition for performing this analysis in the first place --- i.e.~that
the coupling strength is large compared to the surface term. That is to
say, this sum always converges in the case of strong coupling. This is a
sufficient but not necessary condition, in that, conversely, the series
need not necessarily diverge for \(V_{nn} \gtrsim P\).

\begin{figure}[htbp]
  \centering
  \annotate{\includegraphics[trim=.25cm .75cm .25cm .15cm,clip]{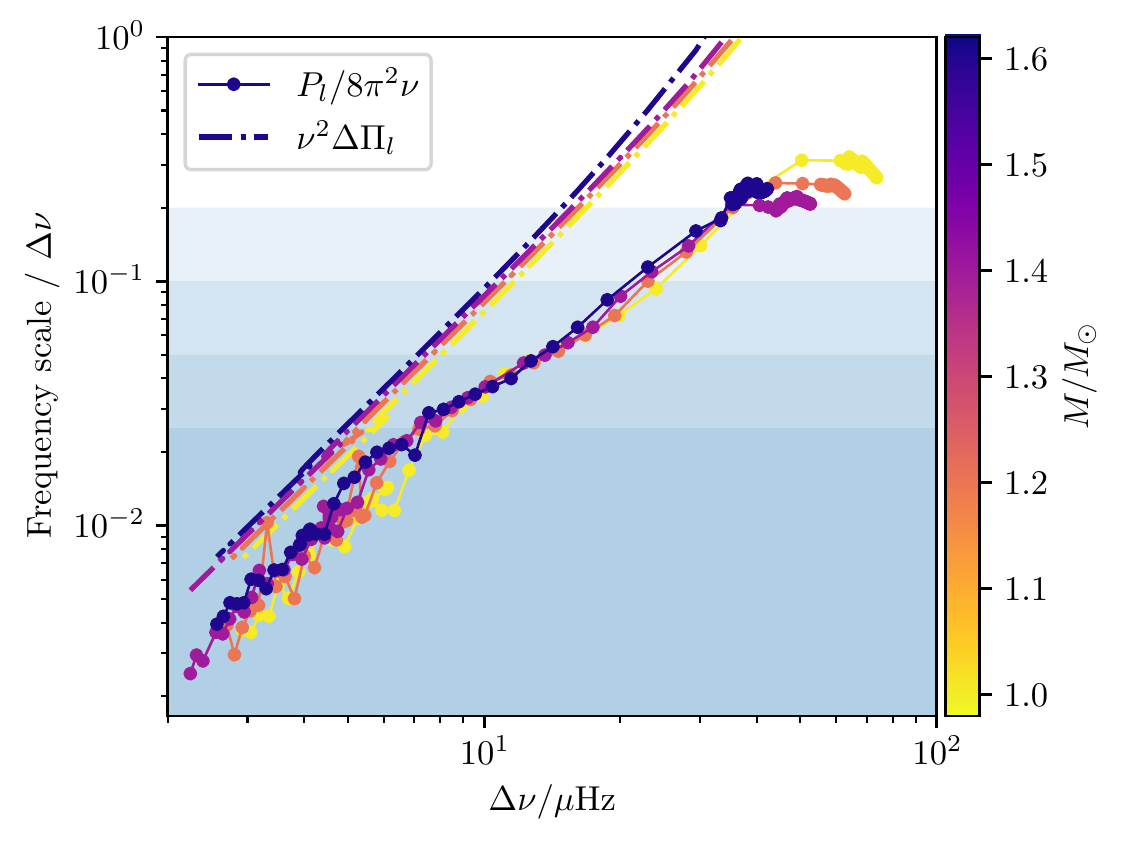}}{
  \node at (.74, .15) {\textbf{(a)}: $l=1$};
  }
  \annotate{\includegraphics[trim=.25cm .25cm .25cm .15cm,clip]{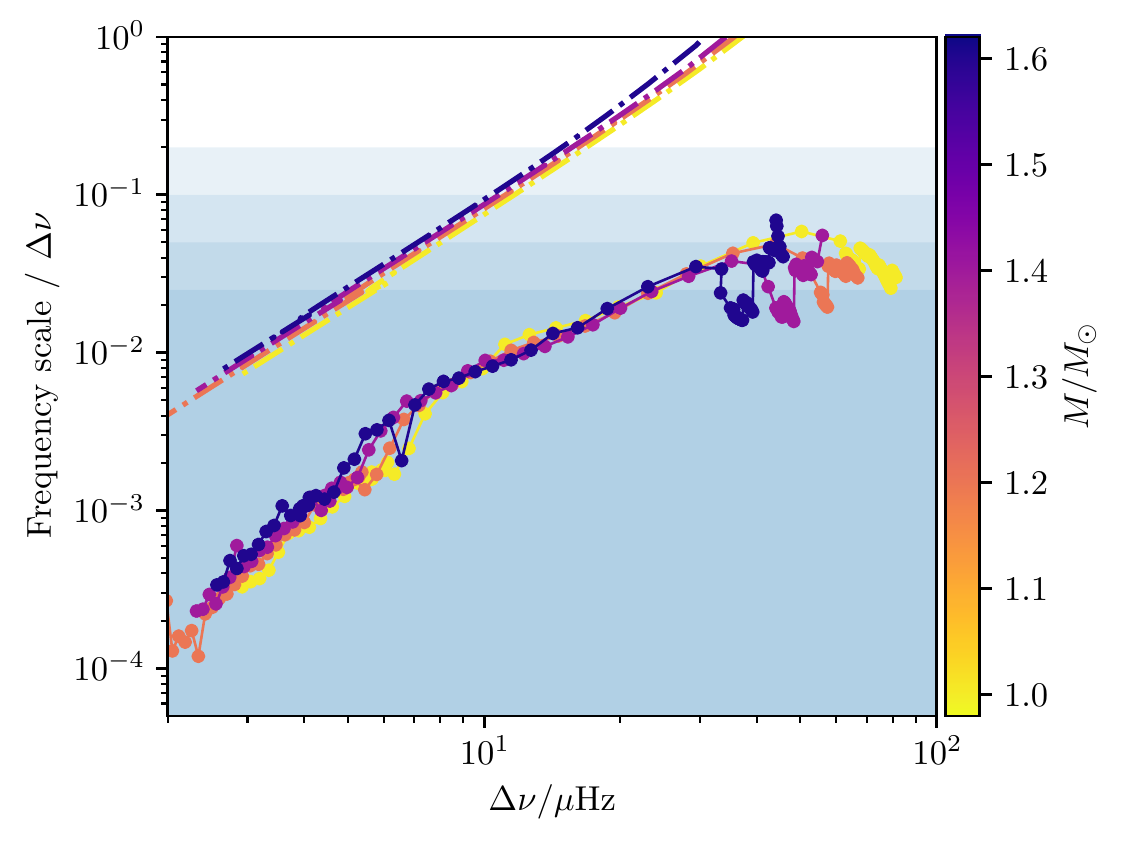}}{
  \node at (.74, .2) {\textbf{(b)}: $l=2$};
  }
  \caption{Evolution of the mixed-mode coupling strength $P/8\pi^2\nu$ (points connected with lines) and g-mode separation $\nu^2\Delta\Pi_l$ (dashed-dotted lines) in the neighbourhood of \numax\ for evolutionary tracks of solar composition. We show these quantities as computed for \textbf{(a)} dipole ($l=1$) and \textbf{(b)} quadrupole ($l=2$) modes. The shaded regions indicate constant fractions of \Dnu\ (see text).\label{fig:regimes}}
\end{figure}

Finally, in the opposite case of weak coupling,
\(V_{i_1i_2} \gg P_{i_1j}\), and we may use \cref{eq:eigk} directly. On
the diagonal, \(B_{ii} = V_{ii}\), while off the diagonal,
\(B_{ij}=P_{ij}\). Moreover, since the coupling is weak, the
perturbative expansion \cref{eq:eigk} is dominated by terms containing
the highest powers of the diagonal elements. For \(k > 2\) we then have
\begin{equation}
\begin{aligned}
\varepsilon_n^{(k)} &\sim (-1)^k\sum_{m \ne n} {\left(V_{nn}\right)^{k-2}{|P_{mn}|}^2 \over \left(\varepsilon^{(0)}_{n} - \varepsilon^{(0)}_{m}\right)^{k-1}},\\
\ c_{nm}^{(k)} &\sim (-1)^{k}{\left(V_{nn}\right)^{k-1}P_{mn} \over \left(\varepsilon^{(0)}_{n} - \varepsilon^{(0)}_{m}\right)^k}.
\end{aligned}
\end{equation} Since the eigenvalues are now given by the bare \(\pi\)
and \(\gamma\) modes, the spacing between them is in principle not
bounded from below (because the bare \(\pi\) and \(\gamma\) eigenvalues
may cross freely over the course of stellar evolution). Conversely, it
is in general bounded from above by \(\min(\Dnu, \nu^2 \Delta\Pi)\). In
the weak coupling regime in particular, we moreover have
\(\nu^2 \Delta\Pi < \Delta\nu\). Accordingly, the perturbation
coefficients are bounded from below as \begin{equation}
  \left|\varepsilon_n^{(k)}\right| \gtrsim {P^2\over |V_{nn}|} \cdot \left(\delta\nu_\text{surf} \over \nu^2 \Delta\Pi\right)^{k-1}.
\end{equation} Note that the direction of the inequality is now
reversed: we have derived that this series \emph{fails to converge} if
\(\delta\nu_\text{surf} > \nu^2 \Delta\Pi\). Again, the converse is not
necessarily true (i.e.~the series may not necessarily converge even if
\(\delta\nu_\text{surf} < \nu^2 \Delta\Pi\)). Comparing values of
\(\Delta\Pi\) and the coupling strength as computed from models reveals
that the two quantities roughly scale with each other over the course of
stellar evolution. Consequently this condition and \cref{eq:bound2} are
mutually exclusive (see our subsequent description of
\cref{fig:regimes}).

There is one final asymptotic regime under which we may once again rely
on perturbation theory even where this series diverges.
\citet{ball_surface_2018}, \ob, and \citet{ong_differential_2021} show
that if \(\Dnu \gg \nu^2 \Delta\Pi \gg P/8\pi^2\nu\), then the density
of \(\pi\)-modes is so high, and the \(\pi\)-\(\gamma\) coupling is so
weak, that the p-dominated mixed mode frequencies which emerge are
well-approximated by those of pure \(\pi\)-modes. In these cases we may
safely treat the effects of the surface term on the observed modes as we
would if they were bare \(\pi\)-modes, and ignore the details of mode
coupling altogether. This is already common practice for treating the
quadrupole modes of evolved red giants.

In summary: the traditional ``variational'' analysis, from which the
\bg~surface term is derived, always holds good when applied to
\(p\)-modes or bare \(\pi\)-modes. However, such analyses cannot always
be applied to mixed modes. We have shown that if the frequency shift
within the bare \(\pi\)-mode subspace is given by
\(\delta\nu_\text{surf}\), then these constructions may only be applied
to mixed modes where it is smaller than the smallest separation between
adjacent modes. This may be variously specified by \(\Dnu\), the
relative coupling strength \(P/8\pi^2\nu\), or the local g-mode spacing
\(\nu^2\Delta\Pi\) at different stages of the evolution of the star off
the main sequence. These bounds follow from quite general
considerations, which also will apply to other contexts where such
variational constructions are typically invoked (e.g.~in the
construction of structural or rotational inversion kernels).

To illustrate these various regimes of approximation, we plot in
\cref{fig:regimes} the relative coupling strength \(P/8\pi^2\nu\) (found
by averaging matrix elements near \(\numax\)), and the local
\(\gamma\)-mode separation \(\nu^2\Delta\Pi\), as computed with respect
to \mesa~models with solar-calibrated \(\amlt\) and \(Y_0\), on
evolutionary tracks with the \gs~element mixture at solar metallicity.
For this purpose we use the same evolutionary tracks as in
\citet{ong_structural_2019}. However, following
\citet{benomar_masses_2012} (who used an approximate parametrisation of
the coupling strength), we expect these quantities to at least
qualitatively be independent of the composition of the stellar models.
We show these quantities as averaged over \(\pi\) and \(\gamma\) modes
within \(4\Dnu\) of \(\numax\), in units of \(\Dnu\). We compute these
quantities for evolutionary tracks with stellar masses between 1 to 1.6
\(M_\odot\), from the onset of mode mixing (i.e.~where the lowest-order
\(\gamma\) mode enters into the frequency range of interest) to past the
RGB bump.

Since the surface term frequency perturbation \(\delta\nu_\text{surf}\)
is typically assumed to be a small fraction of \(\Dnu\), we mark out the
constant multiples \(0.2\), \(0.1\), \(0.05\) and \(0.025\Dnu\) with the
horizontal shaded regions. The different regimes of convergence we have
described can be read off immediately from these diagrams for any choice
of \(\delta\nu_\text{surf}\). For example: with
\(\delta\nu_\text{surf} \sim 0.1\Dnu\) near \numax, we find that
first-order constructions like that of \bg~are only reliable for dipole
modes where \(\Dnu \gtrsim 20~\mu\)Hz, while no perturbative treatment
is valid for \(\Dnu \lesssim 10~\mu\)Hz. For
\(\Delta\nu \lesssim 20~\mu\)Hz, however, we also see that, since the
quadrupole coupling strength is orders of magnitude weaker, we may
neglect mode mixing altogether for quadrupole and higher-degree modes.

\hypertarget{epsilon-matching}{%
\section{\texorpdfstring{\(\epsilon\)-matching}{\textbackslash epsilon-matching}}\label{epsilon-matching}}

While parametrisations like that of \bg~can be expressed globally
(i.e.~as closed-form integrals over the stellar structure), the
so-called \(\epsilon\)-matching algorithm, which we describe below,
operates on the partial-wave phase functions of the pulsation
eigenfunctions, and does not as easily admit such a global description.

For pure \(p\) or \(\pi\) modes, the radial displacement wavefunctions
of partial waves at angular frequency \(\omega\) may be approximated
near the centre of the star with spherical Bessel functions as
\begin{equation}
  \xi_{r, l}(\omega, t) \sim A_l(\omega, t) j_l\left(\omega t - \delta_l(\omega, t)\right),
\end{equation} where \(A_l\) is an inner amplitude function,
\(\delta_l\) is the inner phase function, and
\(t = \int_0^r \mathrm dr/c_s\) is the acoustic radial coordinate. A
similar approximation can be made near the outer boundary of the star,
with corresponding outer amplitude and phase functions \(B_l\) and
\(\alpha_l\). Further self-consistency requirements
\citep{calogero_novel_1963, babikov} permit \(\alpha_l\) and
\(\delta_l\) to be specified uniquely, up to integer multiple of
\(\pi\), by integrating a set of nonlinear ordinary differential
equations starting from the outer and inner boundaries, respectively.
\citet{roxburgh_surface_2015} showed that both \(\alpha_l\) and
\(\delta_l\) tend towards fixed values when evaluated at suitable
matching points \(t_0, 0 < t_0 < T\), sufficiently far from both the
inner and outer boundaries at \(t = 0\) and \(t = T\); their values
there may be treated as functions of frequency alone. When \(\omega\) is
the angular frequency of a normal mode, they must also satisfy the
eigenvalue equation \begin{equation}
  \omega_{nl} T + \left(\alpha_l(\omega_{nl}) - \delta(\omega_{nl})\right) \equiv \omega_{nl} T - \pi \epsilon_l(\omega_{nl}) = n \pi.
\end{equation} Since the stratification in the outer layers of the star
is approximately plane-parallel, the outer phase functions \(\alpha_l\)
do not change significantly with \(l\) at low degree.
\citet{roxburgh_asteroseismic_2016} exploited this property to devise a
procedure for nonparametric diagnosis of the surface term, by comparing
values of \(\epsilon_l\) associated with model and observed frequencies
(whence the name \(\epsilon\)-matching). We generalise the approach in
this section to account for mode mixing, both at first order in the
coupling strength, as well as in more general cases.

Ordinarily, the \(\epsilon\)-matching algorithm acts on observables of
the form \begin{equation}
\mathcal{E}_l(\nu^\text{obs}_{l, n}) = \epsilon^\text{obs}_{l, n} - \epsilon_l^\text{mod}(\nu^\text{obs}_{l, n}),
\end{equation} which we rewrite in terms of frequency differences as
\begin{equation}
\begin{aligned}
\mathcal{E}_l(\nu^\text{obs}_{l, n}) &= \epsilon^\text{obs}_{l, n} - \epsilon_l^\text{mod}(\nu^\text{mod}_{l, n}) + \epsilon_l^\text{mod}(\nu^\text{mod}_{l, n}) - \epsilon_l^\text{mod}(\nu^\text{obs}_{l, n}) \\
&= {\nu^{\text{obs}}_{l, n} - \nu^{\text{mod}}_{l, n} \over \Dnu} + \left(\epsilon_l^\text{mod}(\nu^\text{mod}_{l, n}) - \epsilon_l^\text{mod}(\nu^\text{obs}_{l, n})\right).\label{eq:differences}
\end{aligned}
\end{equation} Following our discussion in the previous section, the
perturbation to the frequencies from the surface term results in a
perturbation to the \(\pi\) subsystem alone, and we assume it to leave
the \(\gamma\)-mode frequencies unchanged. In the presence of mode
mixing, this means (replicating conventional wisdom) that
\(\gamma\)-dominated modes are less strongly affected by the surface
term than are \(\pi\)-dominated ones.

\hypertarget{first-order-construction}{%
\subsection{First-order construction}\label{first-order-construction}}

\label{sec:epsmatching}

Let us first consider a mixed mode such that:

\begin{enumerate}
\def\labelenumi{\roman{enumi}.}
\tightlist
\item
  Only a single \(\pi\) mode is significantly coupled to the
  \(\gamma\)-mode system. The mode eigenfunctions can then be written in
  the form \begin{equation}
    \xi_{\text{mixed},i} = c_{\pi,i} \xi_{\pi,i} + \sum_j c_{\gamma,ij}\xi_{\gamma,j}.
  \end{equation}
\item
  The frequency perturbation from the surface term is such that the
  coefficients \(c_\pi\) and \(c_\gamma\) are not significantly
  affected. This means that the surface term does not modify which
  \(\pi\) and \(\gamma\) modes happen to be in resonance.
\item
  The diagonal elements of the perturbation matrix are sufficient to
  specify the effect of the surface term on the \(\pi\)-mode subspace.
  We write this frequency shift from the surface term as
  \(V_{ii} \equiv -\delta\omega_{\pi,i}^2\).
\end{enumerate}

For a mixed mode satisfying these conditions, retaining terms to leading
order in the perturbation gives us that \begin{equation}
  \delta\omega^2_{\text{mixed},i} \sim \delta\omega_{\pi,i}^2 c_{\pi,i}\left(c_{\pi,i} + \sum_j c_{\gamma,ij}D_{ij}\right) \sim\delta\omega_{\pi,i}^2 / Q_\text{mixed},\label{eq:dw2}
\end{equation} where \(Q_\text{mixed} = I_\text{mixed} / I_{\pi,i}\) is
the ratio between the mode inertia of the mixed mode and that of its
underlying \(\pi\) mode. In the second step we have dropped the overlap
terms \(D_{ij}\). If this frequency perturbation is sufficiently small,
we then have \begin{equation}
  \delta\nu_{\pi,i} \sim Q_\text{mixed}\cdot\delta\nu_\text{mixed}. \label{eq:inertia}
\end{equation} Accordingly, we obtain \begin{equation}
\begin{aligned}
  \mathcal{E}_{\pi,l}(\nu^\text{obs}_{l, n}) &= {\epsilon_\pi^\text{obs}}_{l, n} - \epsilon_{\pi,l}^\text{mod}(\nu^\text{obs}_{l, n})
  \\ &\sim \left(\nu^{\text{obs}}_{l, n} - \nu^{\text{mod}}_{l, n} \over \Dnu\right)Q_{ln} + \left(\epsilon_{\pi,l}^\text{mod}(\nu^\text{mod}_{l, n}) - \epsilon_{\pi,l}^\text{mod}(\nu^\text{obs}_{l, n})\right).\label{eq:modifiedeps}
\end{aligned}
\end{equation} For two stellar structures that differ only in their
outer layers, this can be related to the differences between their outer
partial-wave phase functions, in the same way as
\citet{roxburgh_asteroseismic_2016}: \begin{equation}
  \mathcal{E}_{\pi,l}(\nu^\text{obs}_{l, n}) \sim \alpha^\text{obs}_{\pi,l}(\nu^\text{obs}_{l, n}) - \alpha^\text{mod}_{\pi,l}(\nu^\text{obs}_{l, n}),
\end{equation} which should as usual collapse to a single function of
frequency, \(\mathcal{F}(\nu)\). Since the \(\pi\)-mode isolation
condition is only applied in the interior of the star,
\(\alpha_\pi(\nu) \sim \alpha_p(\nu)\), hence this expression can also
be applied to radial modes using interpolants constructed from the
(technically inequivalent) uncoupled \(p\)-modes.

Therefore, this yields the following modified \(\epsilon\)-matching
procedure:

\begin{enumerate}
\def\labelenumi{\arabic{enumi}.}
\tightlist
\item
  We match model and observed modes pairwise, assuming some reasonable
  mode identification.
\item
  For each mode pair, the quantity \(\mathcal E_{\pi,l,n}\) in
  \cref{eq:modifiedeps} is evaluated. We treat these as samples of some
  function \(\mathcal E_{\pi,l}(\nu_{n,\text{obs}})\) evaluated at the
  observed frequencies.
\item
  The best-fitting model is the one for which these different sets of
  samples collapse to a single function of frequency, such that
  \(\mathcal E_{\pi,l}(\nu) = \mathcal{F}(\nu)\) independently of \(l\).
  As in \citet{roxburgh_asteroseismic_2016}, we characterise this by way
  of a reduced-\(\chi^2\) cost function \begin{equation}
    \chi^2_\epsilon = \min_{M_i, \{\theta_i\}} {1 \over N - M_i - 1} \sum_{l,n}\left(\mathcal{E}_l\left(\nu_{l,n}^{\text{obs}}\right) - \mathcal{F}\left(\nu_{l,n}^{\text{obs}}; \{\theta_i\}\right) \over \sigma_{\mathcal{E}_{ln}}\right)^2,
  \end{equation} where \(M_i\) counts the number of free parameters
  \(\theta_i\) used in the fit to constrain \(\mathcal{F}(\nu)\).
\end{enumerate}

This procedure acts on pairwise frequency differences rather than on the
phase functions, which cannot be directly estimated from the observed
frequencies. At the same time, the explicit dependence on the
\(\pi\)-mode phase function of the fiducial model eliminates the
degeneracy under homology transformations that would otherwise emerge
when considering only frequency differences. This works in much the same
way as described for the original construction of
\citet{roxburgh_asteroseismic_2016}.

Since \(I_{\pi,i} = I_\pi(\nu_i)\) is, for low-degree modes, a
degree-independent function of frequency, \cref{eq:inertia} suggests
that surface term treatments of the form
\(\nu_\pi \mapsto \nu_\pi + f(\nu_\pi)\) can be modified for use with
this restricted class of mixed modes as
\(\nu_i \mapsto \nu_i + [f(\nu_i) I_\pi(\nu_i)] / I_i\), where the
quantity in the square brackets is itself a function of frequency.
Conversely, any such inertia-weighted correction is implicitly also only
applicable to mixed modes satisfying the criteria (i) and (ii) that we
have laid out above, whether or not the perturbation-theory
considerations we have discussed in \autoref{sec:convergence} were
explicitly invoked in its construction.

We now identify several sources of systematic error arising from
approximations made in this construction, and attempt to provide rough
estimates of their relative importance. Notably, we have made much
stronger explicit assumptions in deriving this procedure than are made
for most parametric methods. First, we have assumed that the
coefficients \(c_{i,\pi}^2\) can be reasonably approximated with
\(1/Q\); under unit normalisation, we have \begin{equation}
\begin{aligned}
Q = {I \over I_{\pi,i}} \sim {\oint \mathrm d \Omega |\vec{\xi}(R)_{\pi,i}|^2 \over \oint \mathrm d \Omega |c_{\pi,i}\vec{\xi}(R)_{\pi,i} + \sum_j c_{\gamma,ij}\vec{\xi}(R)_{\gamma,j}|^2} \sim {1 \over c_{\pi, i}^2},
\end{aligned}
\end{equation} where \(I\) is the mode inertia. This expression is exact
only if the \(\gamma\)-mode eigenfunctions vanish at the surface. This
is also implicit in the assumption that the surface term does not affect
the \(\gamma\) modes.

Whereas we might apply mechanical boundary conditions that force the
\(\gamma\)-mode eigenfunctions to vanish at the surface when solving for
\(\gamma\)-modes \citep[as done in][]{ong_semianalytic_2020}, the g-like
components of actual mixed modes are more closely described by those
returned under more standard boundary conditions (e.g., vanishing
Lagrangian pressure perturbation at the surface). The corresponding
inertiae are finite, if large, and decrease with increasing frequency;
equivalently, \(|\vec{\xi}_{\gamma,i}(R)|\) is small, but increases with
frequency. Nonetheless, even then we will expect that
\(\xi_\gamma \sim r^{-l-2}\) within the convective region as we approach
the surface \citep[e.g.][]{pontin_semiconvective_2020}, so we do not
expect this to be significant.

Moreover, in \cref{eq:dw2}, we have further assumed that the overlap
integrals \(D_{ij}\) can be neglected. A priori, we expect this to hold
in the asymptotic limit of high \(n_\gamma\), but in this case we wish
to apply this construction to the regime of isolated avoided crossings,
which involve \(\gamma\)-modes of low radial order. More generally, we
expect these overlap integrals to scale with the coupling strength
\({R_{\pi\gamma,}}_{ij}\) between the \(\pi\) and \(\gamma\) modes. This
coupling strength is largest at low \(n_\gamma\) and low \(l\), since
this is where the \(\pi\)-modes penetrate most deeply, and the
\(\gamma\)-modes decay the most slowly into the convection zone.
Consequently, the errors incurred by this approximation will be largest
in the regime of isolated low-degree avoided crossings ---
unfortunately, precisely where we intend to use this construction.
However, even in this worst case, our subsequent calculations with
stellar models indicate that we should not expect \(D_{ij}\) to be much
larger than \(\approx 10^{-1}\).

A more fundamental assumption going into this construction is item (ii),
which requires that the \(\pi\)-mode mixing coefficient \(c_\pi\) not
differ significantly between each of the observed and model mixed modes.
This is equivalent to demanding that the inertia ratio \(Q\) of the
mixed mode in the actual star is similar to that of the model mixed
mode. Following our discussion in \autoref{sec:perturb}, this assumption
only holds good to first order in the surface perturbation, where the
mixing coefficients are left unchanged. Given the discussion in
\autoref{sec:convergence}, this construction is applicable to either
\(p\)-dominated mixed modes which are far away from resonance with any
\(\gamma\)-modes in both the star and the model, or for mixed modes in
avoided crossings where the frequency shift from the surface term,
\(\delta\nu_\text{surf}\), is much smaller than the coupling strength of
the avoided crossing. Consequently, this particular assumption holds
best in the strong-coupling limit --- i.e.~in the same regime for which
\(D_{ij}\) is largest.

In summary, for this construction to work well, we require that (I)
\(D_{ij} \ll 1\), and that (II) \(\delta \nu^2_\text{surf}\) is small
enough that first-order perturbation theory may be applied. Condition
(I) holds best at high \(n_\gamma\) and high \(l\), but is expected to
incur only minor errors, while (II) holds best in the converse limit of
low \(n_\gamma\) and low \(l\).

\hypertarget{the-case-of-many-pi-modes-per-mixed-mode}{%
\subsection{\texorpdfstring{The case of many \(\pi\)-modes per mixed
mode}{The case of many \textbackslash pi-modes per mixed mode}}\label{the-case-of-many-pi-modes-per-mixed-mode}}

The assumption (i) underlying the above construction is equivalent to
demanding an injective mapping from mixed modes to \(\pi\)-modes, in the
sense that each mixed mode is uniquely associated with a single \(\pi\)
mode. This enters into our final expression for the phase difference
function, \cref{eq:modifiedeps}, via the inertia ratio \(Q\). In the
general case where \begin{equation}
  \xi_{\text{mixed}, k} = \sum_i c_{\pi, ki} \xi_{\pi,i} + \sum_j c_{\gamma,kj}\xi_{\gamma,j},
\end{equation} we instead obtain (again neglecting the
\(\pi\)-\(\gamma\) overlap terms) \begin{equation}
  \delta \omega^2_{\text{mixed}, k} \sim \sum_i \delta\omega^2_{\pi, i} c_{\pi, ki}^2.\label{eq:dw2mixed}
\end{equation} Solving for the pure \(\pi\)-mode frequency perturbations
will require inverting the corresponding mixed-mode squared-coefficient
matrix, which cannot easily be related to traditional quantities like
the mode inertiae. That is, we can no longer relate \(c_{\pi,ki}\) to
\(Q_k\). This approach will strictly only be necessary if the coupling
strength between the \(\pi\) and \(\gamma\) mode cavities is very
strong, so that \(\Dnu \ll P/8\pi^2\nu\). However, we see from
\cref{fig:regimes} (and accompanying discussion) that this is never the
case within the observational regime of interest. Therefore, while a
generalisation of our procedure to permit such cases is possible in
principle, in practice it will never be necessary.

\hypertarget{generalised-cost-function}{%
\subsection{Generalised cost function}\label{generalised-cost-function}}

\label{sec:matrix}

Ultimately, we seek to generalise this procedure to cases where
condition (ii) does not hold, which (per our discussion in
\autoref{sec:convergence}) is necessary for evolved stars where
\(\delta\nu_\text{surf} > \nu^2\Delta\Pi\). In such cases, perturbation
theory cannot be relied upon, and we must seek recourse to explicitly
solving the coupled matrix problem, \cref{eq:HEP}. In general, the
recovery of the perturbed \(\pi\)-mode frequencies from the
surface-perturbed mixed modes is an underdetermined problem where we
cannot rely on the coefficient matrices to remain approximately
constant. Therefore, in these cases we cannot use these modes to
directly constrain the best-fitting phase-difference function
\(\mathcal{F}(\nu)\).

However, from \cref{fig:regimes} we can see that the coupling strength
for quadrupole modes is orders of magnitude weaker than for dipole
modes. Since the classical inner turning points of \(\pi\)-modes at
fixed frequency scale as \(r_t \sim \sqrt{l(l + 1)}\), we expect the
coupling at higher degree to be weaker still. Accordingly, in some cases
we may ignore the effects of mode coupling on the quadrupole (and higher
degree) modes, but not dipole modes. In such cases we may then use the
radial and quadrupole (and potentially higher-degree) modes to first
constrain \(\mathcal{F}(\nu)\) separately, and then compute a cost
function from the dipole modes by performing the appropriate implied
correction on the \(\pi\)-mode subsystem.

Let us suppose that we have already constrained \(\mathcal{F}(\nu)\) in
this manner, using \(N_\epsilon\) out of \(N_\text{tot}\) observed
modes. As in \cref{eq:differences}, if the difference between the two
sets of mode frequencies is only the result of the surface term, then
applying a surface-term correction to the model \(\pi\)-mode frequencies
\(\nu_{\pi,\text{mod}}\) should yield a new set of frequencies
\(\nu_{\pi,\text{surf}}\), which must satisfy \begin{equation}
  {\nu_{\pi,\text{surf}} - \nu_{\pi,\text{mod}} \over \Dnu} + \epsilon^{\text{mod}}_{\pi}\left(\nu_{\pi,\text{mod}}\right) - \epsilon^{\text{mod}}_{\pi}\left(\nu_{\pi,\text{surf}}\right) = \mathcal{F}(\nu_{\pi,\text{surf}}).\label{eq:corr}
\end{equation} This is an implicit equation for
\(\nu_{\pi,\text{surf}}\), which must be solved numerically. If we make
the further assumption that
\(\nu_{\pi,\text{surf}} - \nu_{\pi,\text{mod}}\) is small, we may
approximate \(\nu_{\pi,\text{surf}}\) in closed form by Taylor-expanding
\cref{eq:corr} to first order around \(\nu_\text{mod}\), to obtain
\begin{equation}
  \nu_{\pi,\text{surf}} \sim \nu_{\pi,\text{mod}} + {\mathcal{F}(\nu_{\pi,\text{mod}}) \over {1 \over \Dnu} - {\partial \over \partial \nu}\left.\left(\epsilon_{\pi}^{\text{mod}} + \mathcal{F}\right)\right|_{\nu_{\pi,\text{mod}}}}.
\end{equation} Once a full set of \(\nu_{\pi,\text{surf}}\) is obtained,
we then modify the diagonal elements of the matrix \(\mathbf{L}_\pi\)
associated with the \(\pi\)-mode system of the fiducial model. The
resulting surface-perturbed mixed modes associated with the model may
then be computed by solving \cref{eq:HEP} for the mixed-mode
eigenvalues. These mixed modes yield an auxiliary cost function of the
usual form: \begin{equation}
\chi^2_\text{matrix} = \sum_i \left(\nu_{\text{obs},i} - \nu_{\text{mixed,surf},i}\over \sigma_i\right)^2.
\end{equation} The cost function to be associated with the model is then
given by combining contributions from the \(\epsilon\)-matching proper
and these extra terms: \begin{equation}
  \chi^2_\text{tot} = {1 \over N_\text{tot} - M - 1}\left(\chi^2_\text{matrix} + \left(N_\epsilon - M - 1\right)\chi^2_\epsilon\right).
\end{equation}

\hypertarget{tests-using-stellar-models}{%
\section{Tests using stellar models}\label{tests-using-stellar-models}}

\label{sec:numerics}

We test the constructions in \autoref{epsilon-matching} with simple
injection-recovery test cases, using stellar models constructed with
\mesa~\citep{mesa_paper_1, mesa_paper_2, mesa_paper_4}, with their mode
frequencies and coupling matrices computed with GYRE
\citep{townsend_gyre_2013}. We compute mode frequencies in the interval
\([\numax-\nu_0-\Dnu, \numax+\nu_0]\) such that the upper limit of this
interval is set by the maximum value of the acoustic cutoff frequency in
the atmosphere of the fiducial model. This interval is used for
computing both the mixed modes proper, as well as all of the \(\pi\)
modes necessary to compute the matrices in \cref{eq:HEP}. For the
\(\gamma\)-mode system, we instead compute all modes in the interval
\([\numax-\nu_0-2\Dnu, \max_{r < 0.9R} N/2\pi]\), in order to make the
coupling matrices which appear in \cref{eq:HEP} sufficiently complete
for quantitative use.

\begin{figure}
\centering
\includegraphics{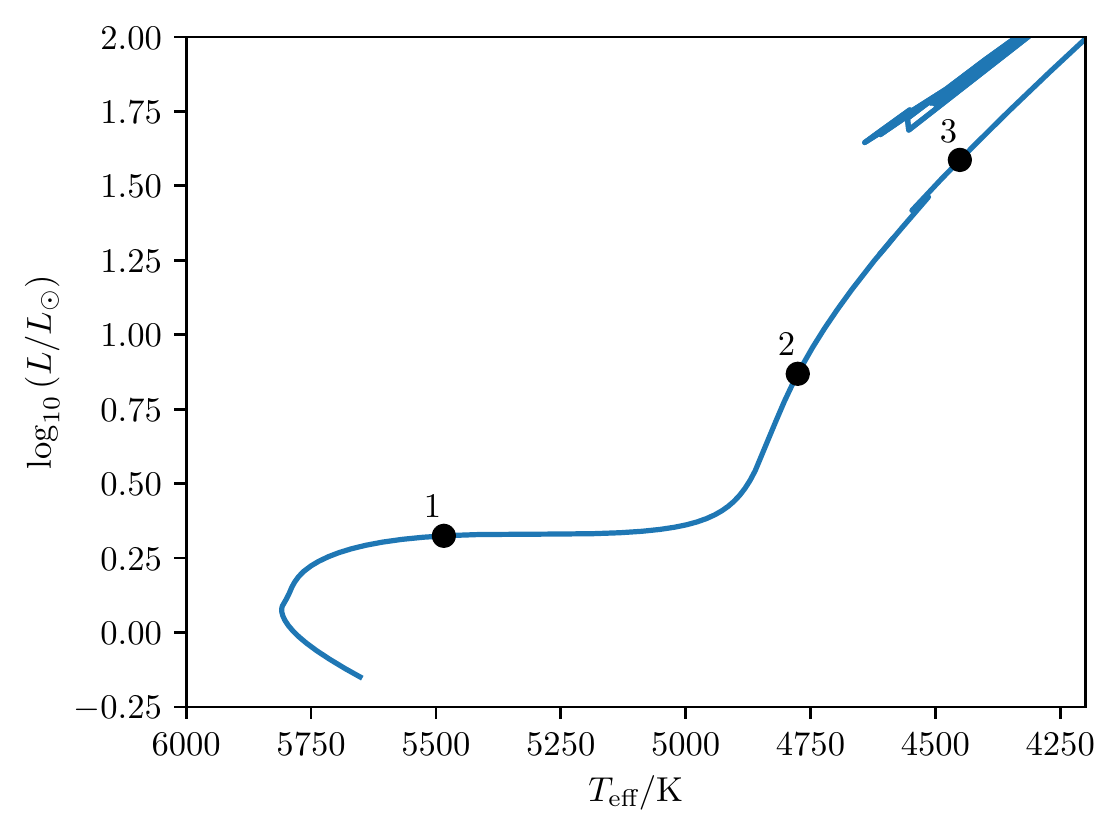}
\caption{\mesa~evolutionary track
\amend{with an initial mass of $1M_\Sun$ and solar composition}, with
the stellar models in subsequent discussion numbered and marked out with
black circles. We examine the effects of the simulated surface term on a
young subgiant crossing the Hertzsprung gap, a young red giant near the
base of the red giant branch, and an older red giant above the RGB
bump.\label{fig:HR}}
\end{figure}

To simulate a surface term, we introduce a structural perturbation to
our chosen fiducial models, and for these perturbed models compute only
the resulting mixed-mode frequencies (i.e.~we do not permit ourselves
access to the underlying decoupled \(\pi\) and \(\gamma\) systems). For
these demonstrations, the precise form of this perturbation is not
important, as long as it is localised to the surface. We choose a
surface perturbation of the form \(P \mapsto P (1 + \Delta)\) and
\(\Gamma_1 \mapsto \Gamma_1 / (1 + \Delta)\) with
\(\Delta = A \exp \left[-(r/R - 1)^2/2\sigma^2\right]\). The size (or
even the sign) of the surface term is essentially a property of the
modelling methodology, and not physically meaningful; cf.~Fig. 2 of
\citet{schmitt_modeling_2015}. For this exercise we limit our attention
to perturbations of a fixed relative size. In our subsequent discussion,
we hold fixed the parameters of this perturbation (choosing
\(A=0.2, \sigma=0.002\)), and examine its effects on three different
stellar models at different points along a solar-calibrated evolutionary
track, whose locations we show on a HR diagram in \cref{fig:HR}.

Since we know a priori that the internal structures of our fiducial and
perturbed models are identical, in principle the resulting cost
functions should all be zero. Therefore, whatever values we do obtain
serve as a relative diagnostic of the systematic error incurred by these
approximations, or elsewhere in the numerical method, relative to some
estimate of the statistical error.

\hypertarget{subgiant}{%
\subsection{Subgiant}\label{subgiant}}

\label{sec:subgiant}

\begin{figure}
\centering
\includegraphics{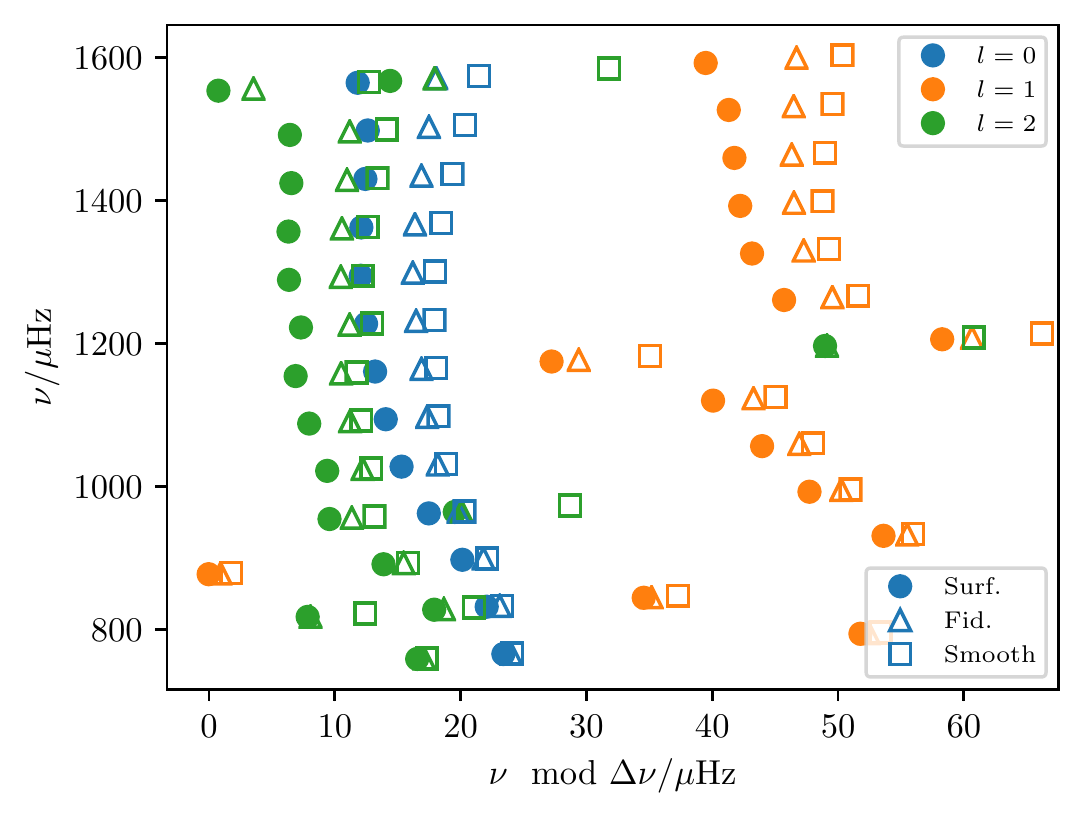}
\caption{Frequency response of a young subgiant \mesa~model to a
perturbation localised to the stellar surface. Mode frequencies from the
fiducial model (\(\lambda = 0\)) are shown with open upright triangles,
and modes from a model with the surface perturbation applied
(\(\lambda = 1\)) are shown with filled circles. Additionally, modes
from a slightly smoothed model (cf.~later discussion in
\autoref{sec:smooth}) are shown with open squares.\label{fig:synthetic}}
\end{figure}

We first examine a young subgiant crossing the Hertzsprung gap (model 1
of \cref{fig:HR}). We compare the frequencies of the perturbed and
fiducial model on an echelle diagram in \cref{fig:synthetic}. This
particular model was chosen because its mode frequencies evince several
instructive features:

\begin{itemize}
\tightlist
\item
  For the dipole modes, we see an on-resonance avoided crossing
  (\(n_\gamma = 1\)) near \(\numax \sim 1200\ \mu\text{Hz}\), and a
  further off-resonance avoided crossing (\(n_\gamma = 2\)) at
  \(\nu \sim 900\ \mu\text{Hz}\). However, the \(\pi\)-\(\gamma\)
  coupling for dipole modes is strong enough for this off-resonance
  avoided crossing to produce significant frequency deviations from the
  asymptotic \(p\)-mode pattern, even several \(\Dnu\) away from the
  underlying \(\gamma\) mode.
\item
  For the quadrupole modes, we see one avoided crossing on resonance
  (\(n_\gamma = 1\)), two more avoided crossings slightly off resonance
  (\(n_\gamma = 3\) and \(4\)), and a \(g\)-dominated mode
  (\(n_\gamma = 2\)) near the dipole avoided crossing. The
  \(\pi\)-\(\gamma\) coupling is visibly much weaker for the quadrupole
  modes than the dipole modes.
\item
  As we expect, the \(g\)-dominated modes are scarcely affected by the
  surface perturbation.
\end{itemize}

\hypertarget{epsilon-matching-1}{%
\subsubsection{\texorpdfstring{\(\epsilon\)-matching}{\textbackslash epsilon-matching}}\label{epsilon-matching-1}}

\begin{figure}
\centering
\includegraphics{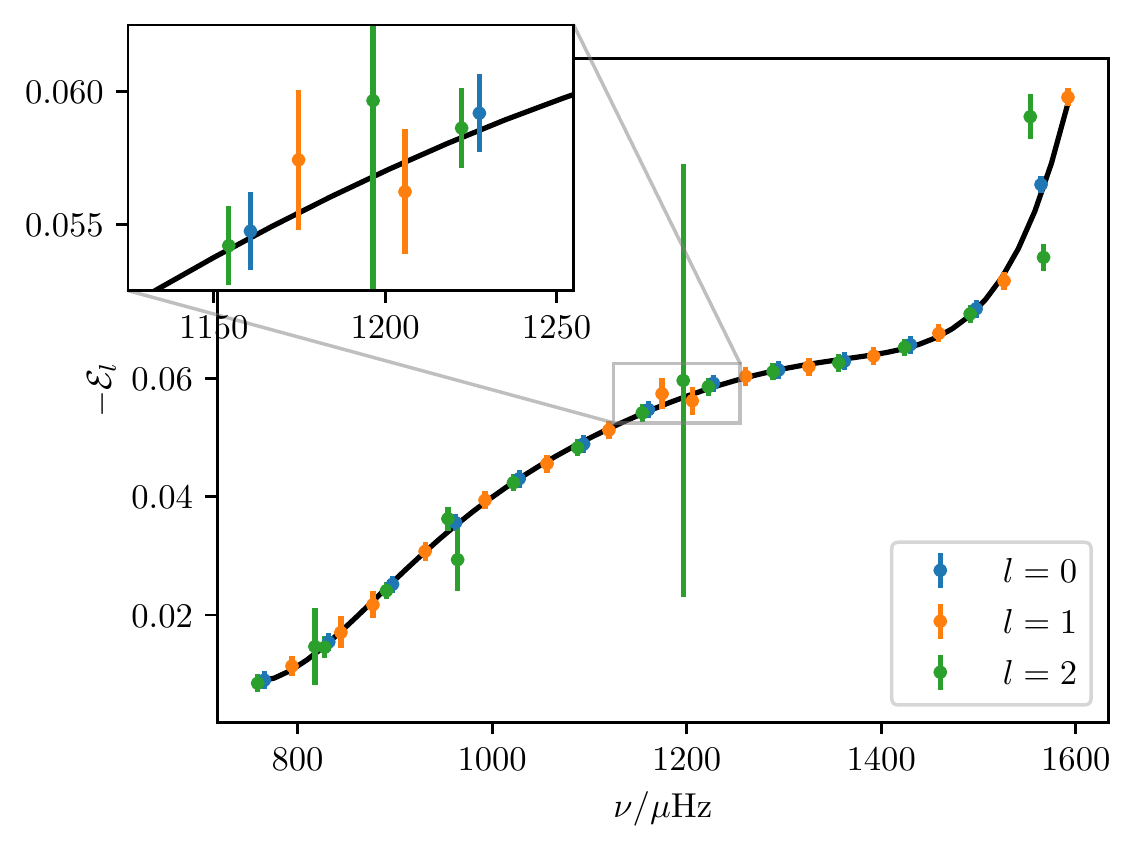}
\caption{Results of applying the modified \(\epsilon\)-matching
procedure to the surface-perturbed subgiant model, shown in
\cref{fig:synthetic}. We show computed values of \(\mathcal{E}_l\) with
points, and a fitted Chebyshev polynomial with the black curve. Isolated
avoided crossings result in a characteristic pattern of residuals that
are first low, then high, on either side of the underlying resonant
modes (as can be seen for dipole modes in inset
panel).\label{fig:synthetictest}}
\end{figure}

We now apply the \(\epsilon\)-matching construction we derived in
\autoref{sec:epsmatching}. For the sake of demonstration, we adopt a
constant value of the mode frequency measurement error of
\(\sigma_\nu = 0.1\ \mu\text{Hz}\), which is comparable to results
returned from \emph{Kepler} \citep[e.g.][]{li_frequencies_2020}. In
practice we should also expect these to increase away from \numax, as
well as for \(g\)-dominated (and so lower-SNR) modes; we do not
explicitly account for these other observational effects in this test.
We show the resulting values of \(\mathcal{E}_l\), and a fitted
Chebyshev polynomial, in \cref{fig:synthetictest}.

We see in \cref{fig:synthetictest} that the \(p\)-dominated modes behave
as we would expect: the residuals to a fit against a single
\(l\)-independent function are dominated by mixed modes near the
isolated avoided crossings. Moreover, we see from our definition in
\cref{eq:modifiedeps} that computing \(\mathcal{E}_l\) from
\(g\)-dominated modes results in propagated measurement errors that are
significantly inflated by their inertia ratios \(Q\); consequently they
do not contribute substantially to the fit.

For our test case, we see that these systematic errors together lead to
a reduced \(\chi^2\) statistic of \(\chi^2_\epsilon = 1.5\), where \(M\)
is the order of the best-fitting Chebyshev polynomial. We interpret this
as a diagnostic of the amount of systematic error incurred by our
construction. Most of this can be attributed to the quadrupole modes
near resonance, for which condition (II) appears not to hold (in
particular for the \(n_\gamma=1\) avoided crossing at high frequencies)
for this particular surface perturbation. This is in turn because the
quadrupole coupling strength is of comparable size to the surface term
perturbation, rendering a first-order construction insufficient; cf.~the
discussion in \cref{fig:regimes}. Excluding these modes from the
computation yields \(\chi^2_\epsilon = 0.13\), indicating that the
first-order approximation works very well for the other modes. On the
other hand, the coupling strength for dipole modes is an order of
magnitude larger, and so the resulting systematic error in
\(\mathcal{E}_1\) is smaller than our adopted trial measurement error,
even near the on-resonance \(n_\gamma=1\) dipole avoided crossing (inset
panel of \cref{fig:synthetictest}).

\cref{fig:synthetictest} shows a characteristic pattern in the residuals
of \(\mathcal{E}_l\) near each avoided crossing, where the values
computed with \cref{eq:modifiedeps} are first lower, then higher, than
would be consistent with a single function of frequency. To leading
order, this is the direct result of neglecting the overlap terms
\(D_{ij}\) in \autoref{sec:epsmatching}. To see why this is so, let us
consider the restricted example of a single \(\pi\) mode coupling to a
single \(\gamma\) mode. This can be described as a perturbed eigenvalue
problem of the form \begin{equation}
  \begin{bmatrix}
  -\omega_p^2 - \lambda V & \alpha - D \omega_\pi^2 \\ \alpha - D \omega_\pi^2 & -\omega_g^2
  \end{bmatrix} \begin{bmatrix} c_p \\ c_g \end{bmatrix} = -\omega^2 \begin{bmatrix}
  1 & D \\ D & 1
  \end{bmatrix} \begin{bmatrix} c_p \\ c_g \end{bmatrix},
\end{equation} with eigenvalues \(\omega^2_+, \omega^2_-\), where \(D\)
is the volume overlap integral defined in \cref{eq:overlap}, and
\(\alpha\) is the \(\pi\)-\(\gamma\) coupling strength. Without loss of
generality, we choose a normalisation of the eigenfunctions so that both
are positive. Moreover, at an avoided crossing, the two modes are in
resonance, so we set
\(\omega_p \sim \omega_g \sim \omega_\pi \sim \omega_0\). Here \(V\)
represents the (small) frequency perturbation from the surface term,
acting only on the \(p\)-mode subsystem, which is ``turned on'' as
\(\lambda\) goes from 0 to 1.

Since this is a \(2\times2\) problem, we can analytically evaluate the
eigensystem of this eigenvalue problem at resonance. For the unperturbed
problem (\(\lambda \to 0\)) we find that \begin{equation}
  \omega^2_\pm = \omega_0^2 \pm {\alpha \over 1 \mp D};\quad\ c_{g,\pm} = \mp c_p.
\end{equation} Let us first consider the result of neglecting \(D\) in
our procedure. The frequency perturbations from the surface term can be
found, to leading order, from perturbation theory with respect to this
mixed mode basis using \cref{eq:normalperturb}, as \begin{equation}
  \delta \omega^2_\pm = \lambda V c_p(c_p + D c_{g,\pm}) + \mathcal{O}(\lambda^2) \sim \lambda c_p^2 (1 \mp D).
\end{equation} But our construction assumes
\(\delta\omega^2 \sim \lambda V c_p^2\) to first order --- i.e.~we have
neglected the overlap terms in \cref{eq:dw2} --- so per
\cref{eq:modifiedeps} we find that this causes us to systematically
underestimate \(\mathcal{E}_l(\nu_-)\) and overestimate
\(\mathcal{E}_l(\nu_+)\) for each isolated avoided crossing.

\hypertarget{sensitivity-to-interior-differences}{%
\subsubsection{Sensitivity to interior
differences}\label{sensitivity-to-interior-differences}}

\label{sec:smooth}

We now examine the ability of this procedure to discriminate between
stars of differing interior structures. For this purpose, we again
compute the eigensystems, of both mixed and \(\pi\) modes, of a model
which is otherwise identical to the fiducial model, except for a very
small amount of smoothing applied to the pressure and density gradients:
3 iterations of a box filter, 2 mesh points wide. We do this to ensure
that the overall acoustic structure is only very slightly changed
compared to the fiducial model.

We show the resulting mixed modes with open squares in
\cref{fig:synthetic}. The effect of this smoothing on the \(p\)-modes is
small and similar to that of a surface term: this can be seen from its
effect on the radial modes. Unlike the surface term, this smoothing also
changes the avoided-crossing pattern, since the frequencies of the
lowest-order \(\gamma\) modes are extremely sensitive to the
Brunt-Väisälä frequency profile. In particular, the \(n_\gamma=1\) and
\(n_\gamma=3\) quadrupole avoided crossings have now gone significantly
off resonance.

\begin{figure}[htbp]
  \centering
  \annotate{\includegraphics[height=2.45in,trim=.15cm .75cm .25cm .15cm,clip]{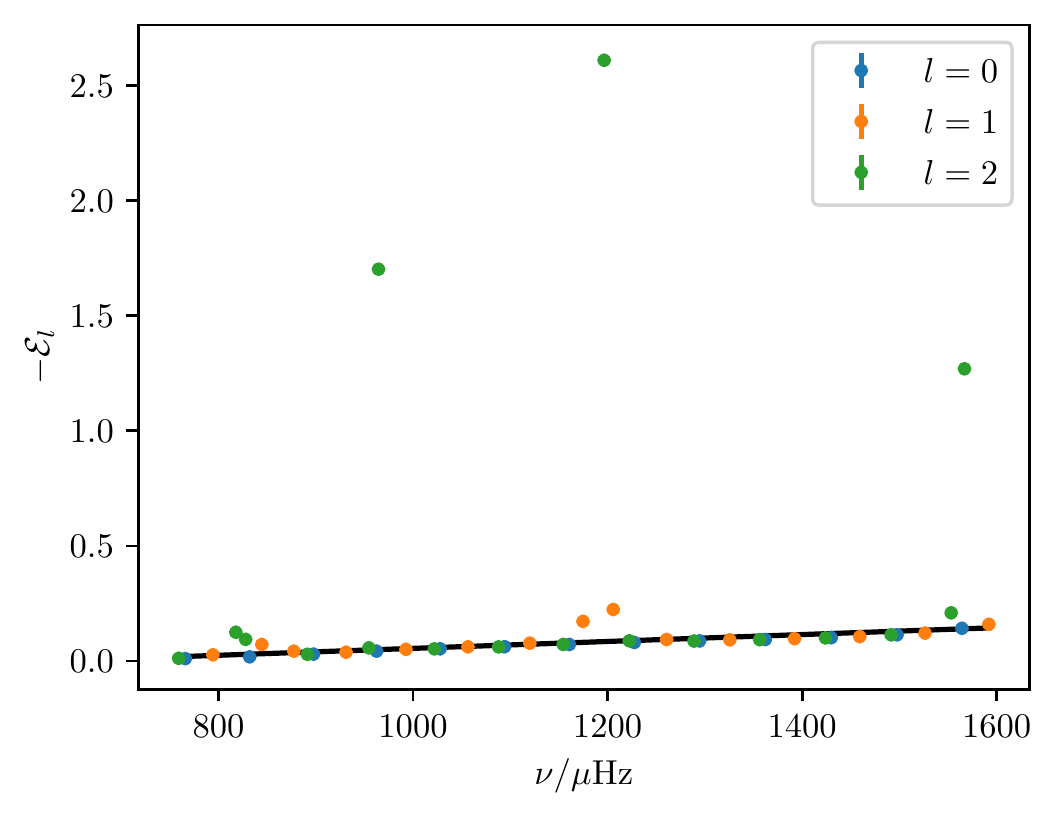}}{
  \node at (.25, .9) {\textbf{(a)}: $\chi^2_\epsilon \sim 1200$};
  }
  \annotate{\includegraphics[height=2.45in,trim=.15cm .25cm .25cm .15cm,clip]{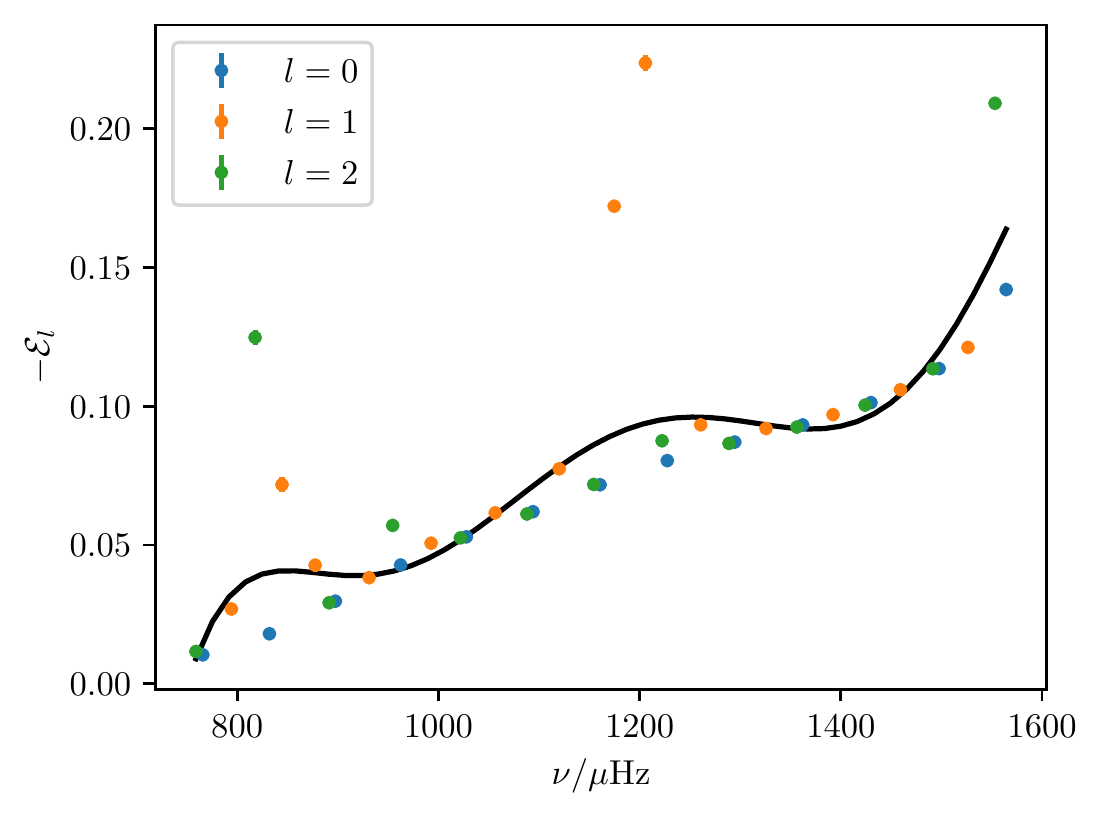}}{
  \node at (.85, .22) {\textbf{(b)}: $\chi^2_\epsilon \sim 209$};
  }
  \caption{Results of applying the modified $\epsilon$-matching procedure to a slightly smoothed model. Values of $\mathcal{E}_l$ and a fitted Chebyshev polynomial are shown with the procedure applied to all modes in \textbf{(a)}, and with the set of modes limited to those with $Q < 2$ in \textbf{(b)}.}
  \label{fig:smooth}
\end{figure}

Applying our \(\epsilon\)-matching procedure to this smoothed model, we
find that the changes to the underlying \(\gamma\)-modes lead to values
of \(\mathcal{E}_l\) that are very large (\cref{fig:smooth}), since they
are heavily upweighted by the inertia ratios \(Q\). These shifts cannot,
and should not, be explicable by any surface term correction. Moreover,
we find that the residuals of the mixed modes near the on-resonance
avoided crossings no longer exhibit the characteristic jagged morphology
we described in the preceding section. Even excluding the most
\(g\)-dominated modes from the cost function of this procedure, by
including only modes with \(Q<2\) in the cost function, we still obtain
a reduced \(\chi^2\) of more than 200. We conclude that this technique
is indeed able to diagnose even minor differences in interior
structures.

\hypertarget{young-red-giant}{%
\subsection{Young red giant}\label{young-red-giant}}

\begin{figure}
\centering
  \annotate{\includegraphics[height=2.45in,trim=.15cm .25cm .25cm .15cm,clip]{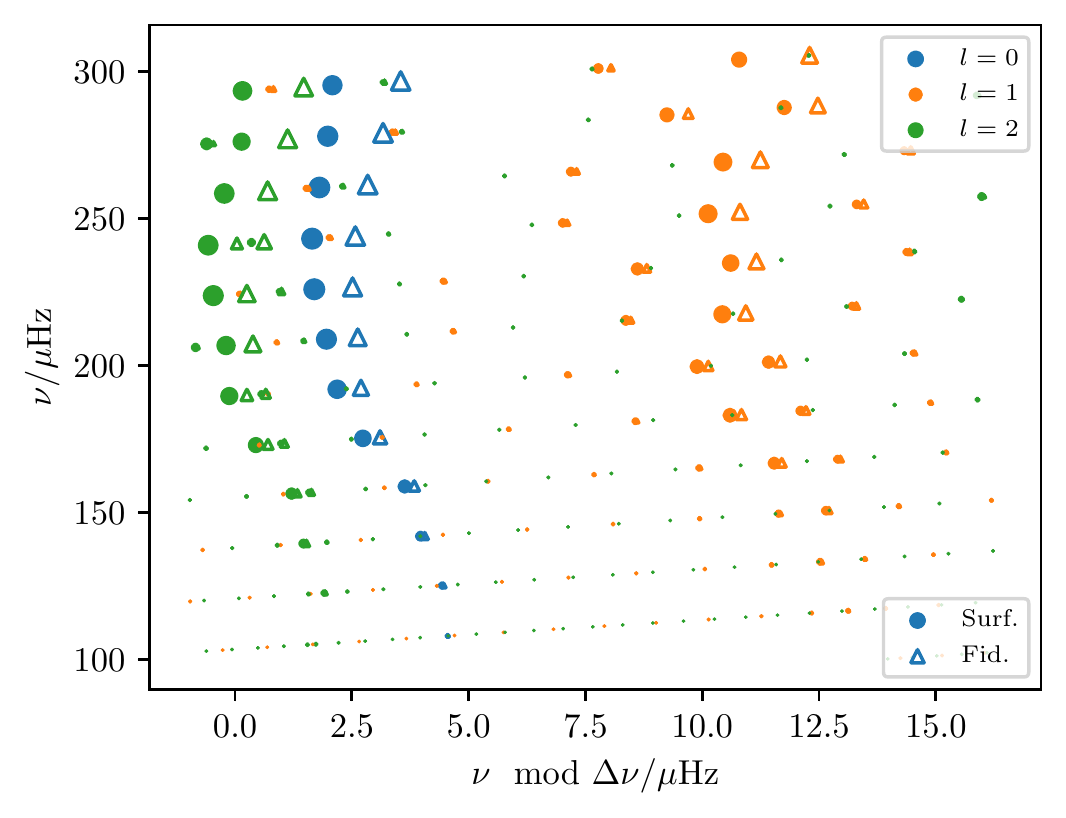}}{
  \node at (.17, .9) {\textbf{(a)}};
  }
  \annotate{\includegraphics[height=2.45in,trim=.15cm .25cm .25cm .15cm,clip]{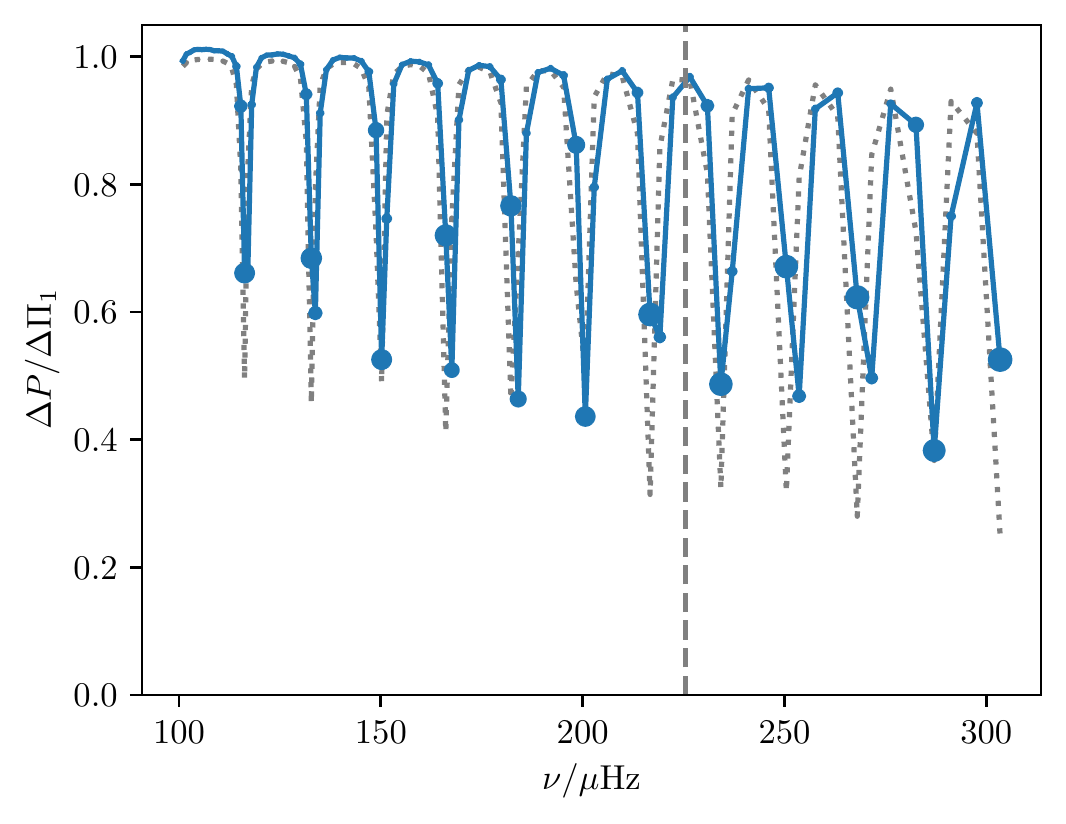}}{
  \node at (.90, .22) {\textbf{(b)}};
  }
\caption{\textbf{(a)}: Fiducial and perturbed frequencies of a young red giant model responding to a perturbation localised to the stellar surface. Markers and colours have the same meaning as in \cref{fig:synthetic}. Points are sized by mode inertiae to indicate whether the modes shown are p- or g-dominated. \textbf{(b)}: Local period differences between adjacent dipole modes as a function of frequency, showing \numax\ with the vertical dashed line. We show the $g$-mode inertia ratio $\zeta \sim 1 - c_\pi^2$ with the grey dotted line in the background.\label{fig:synthetic2}}
\end{figure}

We next apply the same methodology to a slightly more evolved star
(model 2 of \cref{fig:HR}). For this test we have used a young red giant
(near the base of the red giant branch, but before the RGB bump) with
\(\Delta\nu = 17.5~\mu\)Hz. We show its mode frequencies, as well as
those of a model with the surface perturbation applied, in
\cref{fig:synthetic2}a. Of particular note here are a dipole avoided
crossing at \(\sim 290\ \mu\)Hz, and a quadrupole avoided crossing at
\(\sim 240\ \mu\)Hz, in the mode frequencies of the fiducial model.

Whereas the mixed modes for the subgiant shown in \cref{fig:synthetic}
feature the coupling of many \(\pi\)-modes to few \(\gamma\)-modes, the
mixed modes in this case are the result of the opposite scenario (few
\(\pi\)-modes to many \(\gamma\)-modes). From \cref{fig:regimes}, we see
that this model lies close to the limits of validity of first-order
perturbative constructions where the surface term is small
(e.g.~\(\delta\nu_\text{surf}\sim 0.1\Dnu\)) for quadrupole modes, and
outside the formal radius of convergence for dipole modes. We also show
in \cref{fig:synthetic2}b the absolute first differences of the
oscillation periods of the dipole modes, normalised by the asymptotic
period spacing \(\Delta\Pi_1\). When the g-modes are much denser than
the p-modes, this quantity known to approximate the quantity
\(\zeta \sim 1 - c_\pi^2\), which takes values close to 1, except near
the p-modes. We show \(\zeta\) with the grey dotted line in the
background. However, we can see that (particularly above \numax) the
g-modes are sparse enough that this is a poor approximation. Thus, this
red giant is not yet \amend{so} evolved that we may ignore mode mixing
altogether, and rely solely on \(\pi\)-mode computations. In general,
since the size of the surface-term perturbation increases with
frequency, we will expect our approximations to work less well at higher
frequencies than at lower frequencies.

\begin{figure}
\centering
\includegraphics{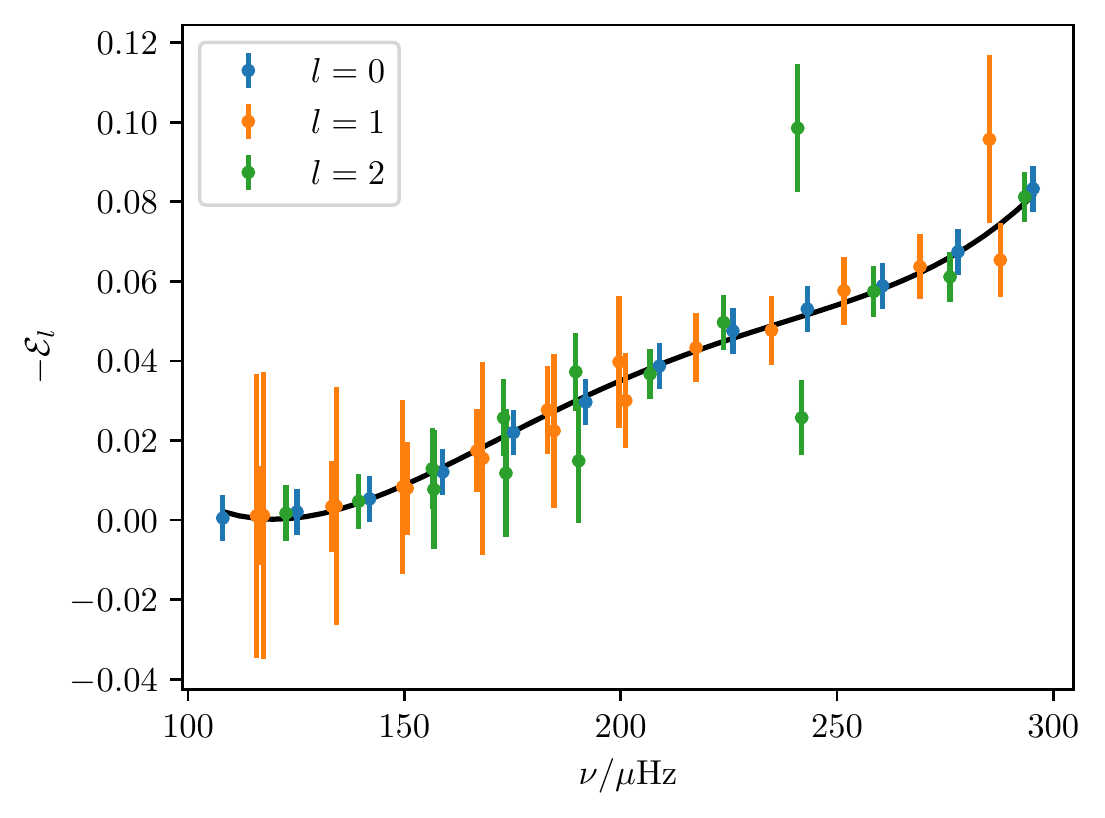}
\caption{Result of applying our first-order construction to the
perturbed model shown in \cref{fig:synthetic2}. Symbols and lines have
the same meanings as in \cref{fig:synthetictest}.
\label{fig:synthetictest2}}
\end{figure}

We first apply our first-order construction from
\autoref{sec:epsmatching}, limiting ourselves to only including modes
where \(Q < 5\). Again we assume a constant frequency uncertainty of
\(0.1~\mu\)Hz. We show the resulting values of \(\mathcal{E}_l\) and the
corresponding best-fitting Chebyshev polynomial in
\cref{fig:synthetictest2}.

The first-order procedure yields a reduced \(\chi^2\) statistic of
\(\chi^2_\epsilon = 0.55\). As in \autoref{sec:subgiant}, most of the
deviation from a single function \(\mathcal{F}(\nu)\) arises in the
neighbourhood of avoided crossings, where the same jagged morphology
emerges for the residuals, although the deviations are more pronounced
here than previously. This is to be expected given the overall weaker
coupling. In particular, we can see in \cref{fig:synthetic2} that the
surface term is large enough, relative to the coupling strength, that
the \(\pi\) mode participating in the quadrupole avoided crossing in the
fiducial frequencies at \(\sim240\ \mu\)Hz has been moved off resonance
with the underlying \(\gamma\) mode. Once again, excluding this
highest-frequency avoided quadrupole avoided crossing from consideration
yields a significantly reduced amount of systematic error, with
\(\chi^2_\epsilon=0.16\).

\begin{figure}[htbp]
  \centering
  \annotate{\includegraphics[height=2.56in,trim=.25cm .25cm .25cm .15cm,clip]{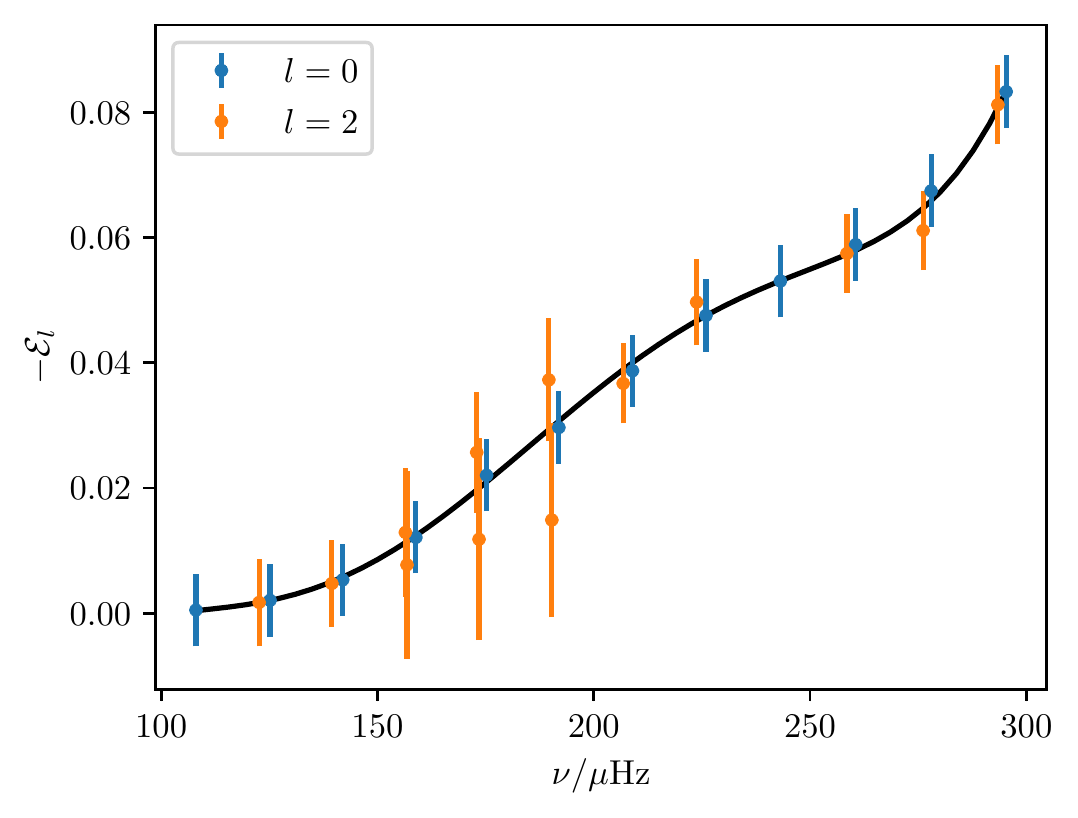}}{
  \node at (.85, .2) {\textbf{(a)}: $\chi^2_\epsilon=0.17$};
  }
  \annotate{\includegraphics[height=2.56in,trim=.25cm .25cm .25cm .15cm,clip]{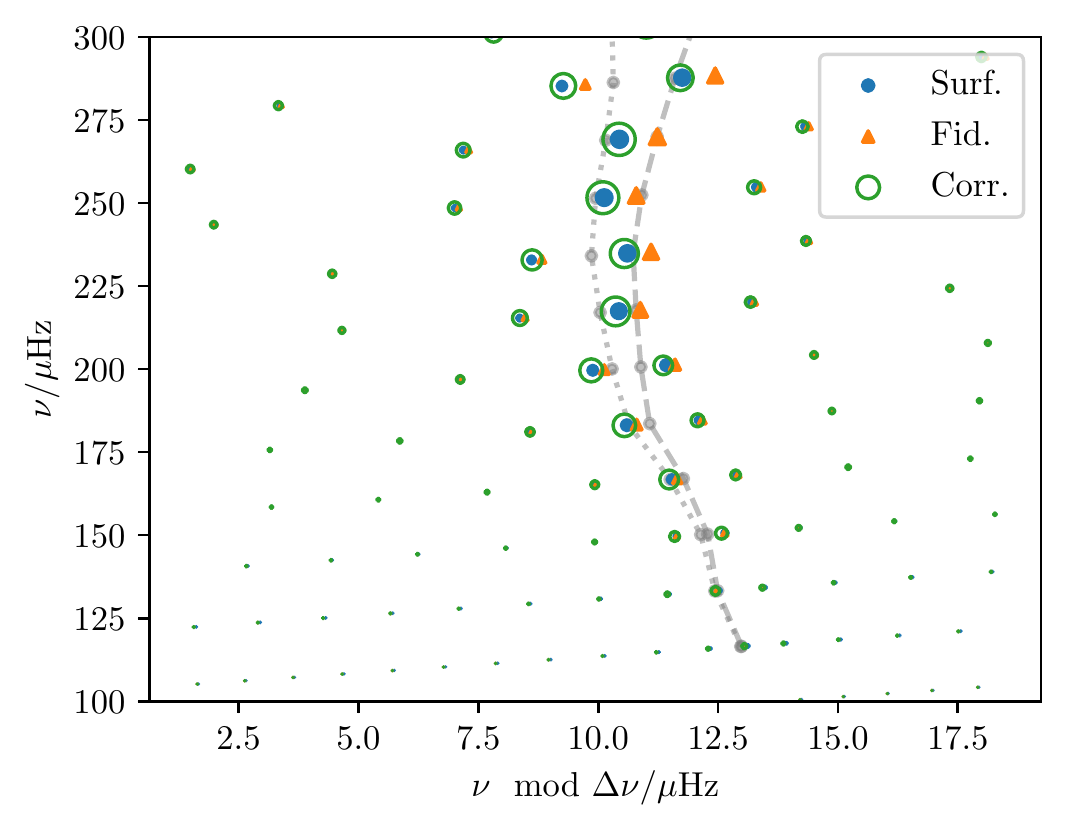}}{
  \node at (.85, .2) {\textbf{(b)}: $\chi^2_\text{tot}=0.21$};
  }
  \caption{Illustration of the procedure described in \autoref{sec:matrix}, as applied to our young red giant model. \textbf{(a)} Modified $\epsilon$-matching construction applied to only radial and quadrupole modes. \textbf{(b)} Dipole modes from both the surface-perturbed model (blue circles) and from the fiducial model, both without (orange triangles) and with (green open circles) the application of the surface correction implied by \cref{eq:corr}. Points are sized by the inverse mode inertia. We also show the underlying $\pi$-mode frequencies without (dashed grey line) and with (dotted grey line) this same correction.}
  \label{fig:corrmatrix1}
\end{figure}

For comparison, we also apply the generalised correction described in
\autoref{sec:matrix}, fully accounting for the coupling between \(\pi\)
and \(\gamma\) modes, which we illustrate in \cref{fig:corrmatrix1}. In
\cref{fig:corrmatrix1}a, we show the results of applying the same
modified \(\epsilon\)-matching procedure, but restricted to only radial
and quadrupole modes. This yields a best-fitting function
\(\mathcal{F}(\nu)\), which we used to generated corrected \(\pi\)-mode
frequencies from those of the fiducial model via \cref{eq:corr}. These
are then used to generate a full set of mixed modes by way of the
prescription in \autoref{sec:matrix}, which we show in
\cref{fig:corrmatrix1}b. In addition to mixed-mode frequencies, we also
show the underlying \(\pi\)-modes of the fiducial model with and without
the same correction being applied. The overall cost function from the
combined procedure is \(\chi^2_\text{tot} = 0.21\).

From \cref{fig:corrmatrix1}b, it is apparent that even where mixed modes
in the fiducial model lie close to their underlying \(\pi\) modes, the
corresponding corrected mixed mode frequencies may not necessarily lie
close to the corrected \(\pi\) mode frequencies. This is unlike the
behaviour of inertia-weighted corrections, which do not explicitly
account for the locations of nearby \(\gamma\) modes. The matrix
construction appears to do an adequate job of the correction, with a
reduced cost function of
\(\chi^2_\text{matrix} / (N_\text{tot} - N_\epsilon - 1)=0.29\)
(limiting ourselves again to only modes with \(Q<5\)).

\hypertarget{evolved-red-giant}{%
\subsection{Evolved red giant}\label{evolved-red-giant}}

\begin{figure}
\centering
  \annotate{\includegraphics[height=2.45in,trim=.15cm .25cm .25cm .15cm,clip]{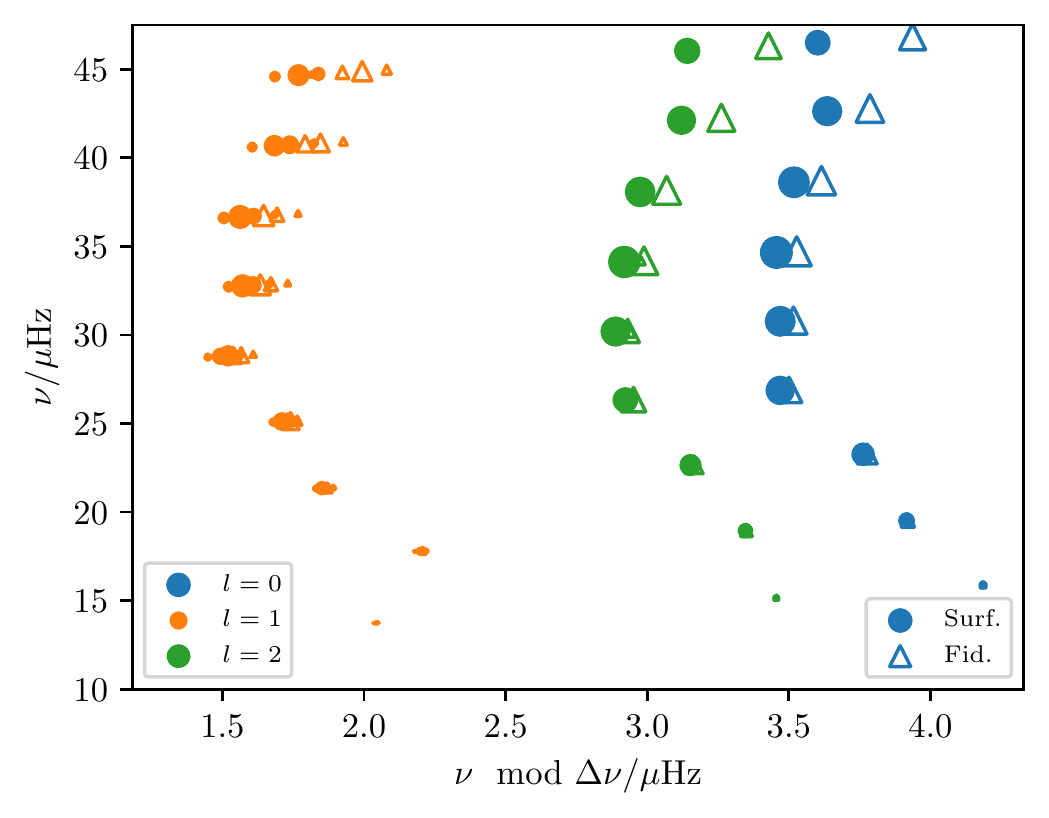}}{
  \node at (.17, .9) {\textbf{(a)}};
  }
  \annotate{\includegraphics[height=2.45in,trim=.15cm .25cm .25cm .15cm,clip]{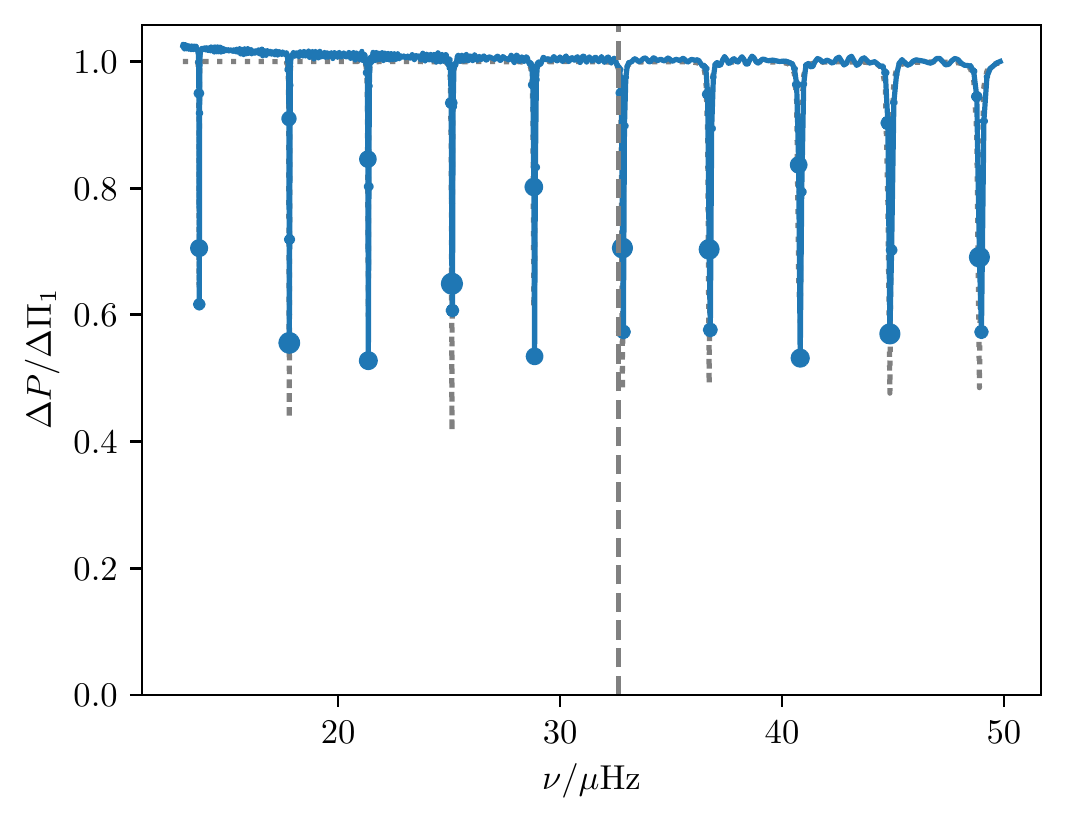}}{
  \node at (.90, .22) {\textbf{(b)}};
  }
\caption{\textbf{(a)}: Fiducial and perturbed frequencies of a evolved red giant model responding to a perturbation localised to the stellar surface. Markers and colours have the same meaning as in \cref{fig:synthetic,fig:synthetic2}. We show only modes with $Q < 25$. Points are sized by mode inertiae to indicate whether the modes shown are p- or g-dominated. \textbf{(b)}: Local period differences between adjacent dipole modes as a function of frequency, showing \numax\ with the vertical dashed line. Again, we show the quantity $\zeta \sim 1 - c_\pi^2$ with the grey dotted line in the background.\label{fig:synthetic3}}
\end{figure}

Finally, we consider these same methods as applied to a still more
evolved red giant (\(\Dnu=3.9\ \mu\)Hz; model 3 of \cref{fig:HR}), with
the same surface perturbation applied. We show the echelle diagram of
its mode frequencies in \cref{fig:synthetic3}a.

Generally speaking, the frequency measurement errors of these evolved
stars are smaller than for subgiants (partly owing to the longer mode
lifetimes). Consequently, for this exercise we adopt a measurement error
of \(0.025\ \mu\)Hz. This time, the size of the surface term
perturbation is comparable to the local \(g\)-mode spacing. Accordingly,
following our discussion in \autoref{sec:convergence}, we may opt to
rely entirely on the \(\pi\)-mode system of our fiducial model, and
eschew the computation of mixed modes altogether.

\begin{figure*}[htbp]
  \centering
  \centering
  \annotate{\includegraphics[height=2.56in,trim=.25cm .25cm .25cm .15cm,clip]{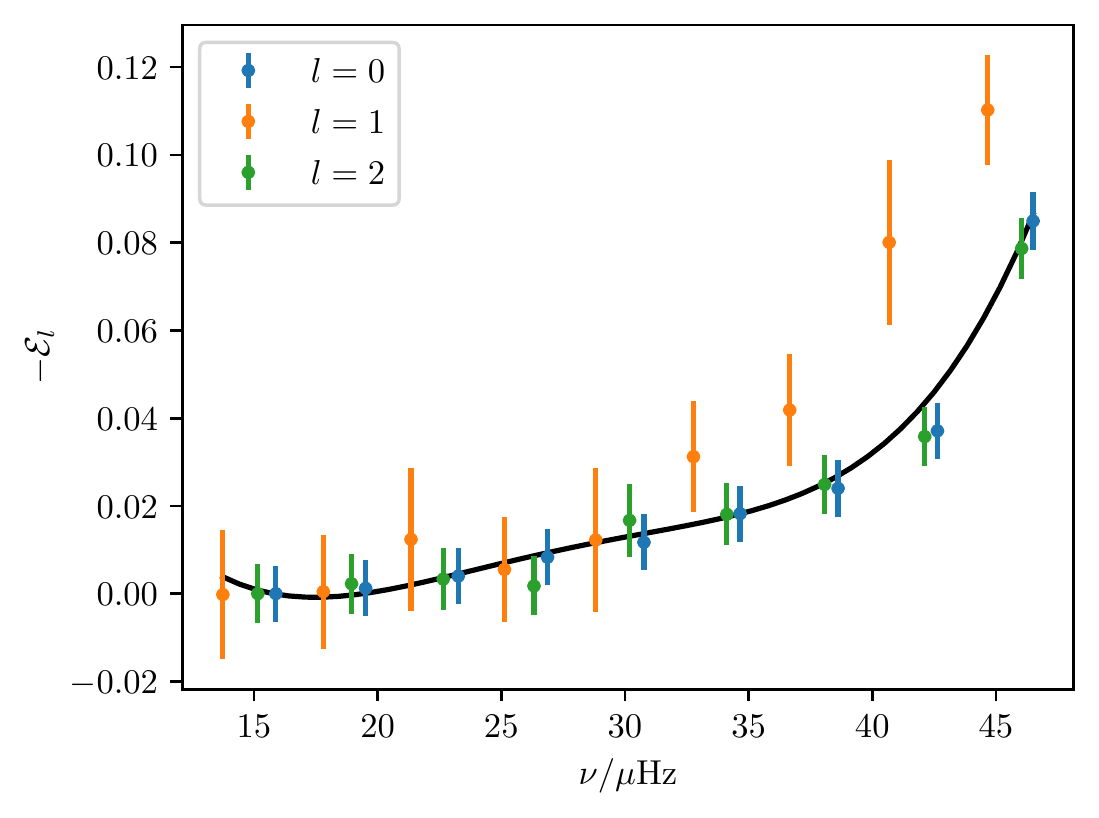}}{
  \node at (.85, .2) {\textbf{(a)}: $\chi^2_\epsilon=1.32$};
  }
  \annotate{\includegraphics[height=2.56in,trim=.25cm .25cm .25cm .15cm,clip]{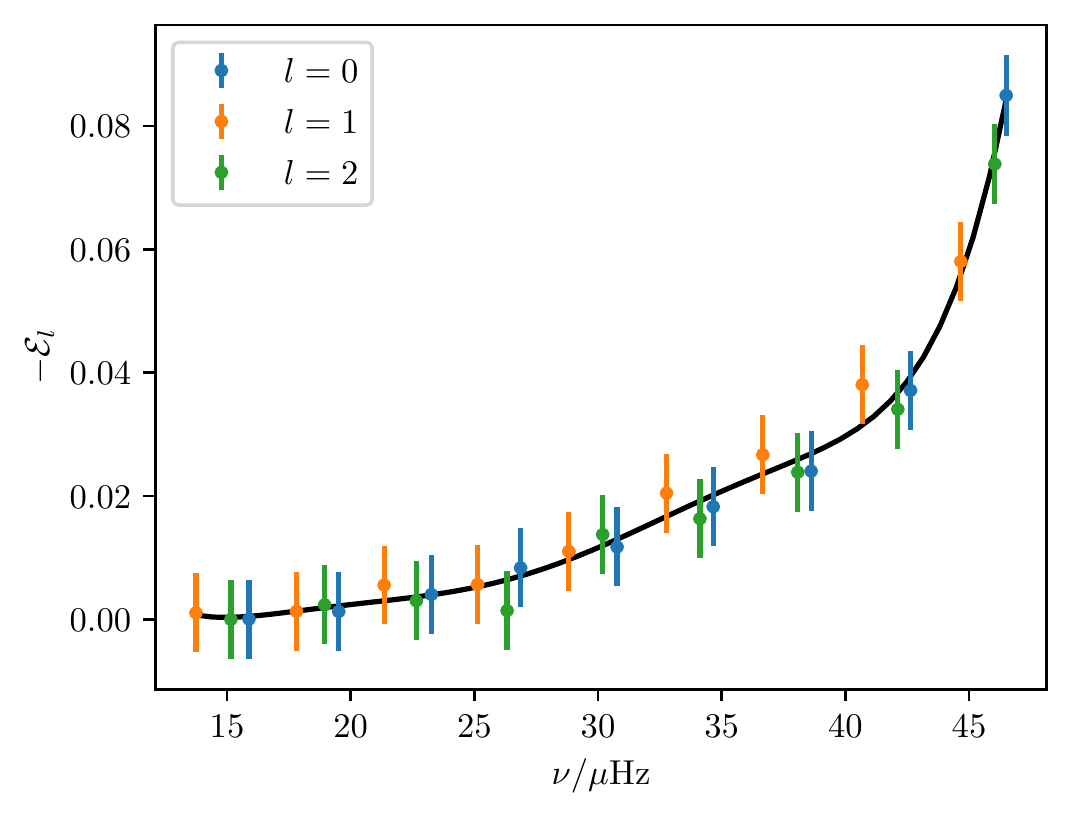}}{
  \node at (.85, .2) {\textbf{(b)}: $\chi^2_\epsilon=0.19$};
  }
  \annotate{\includegraphics[height=2.56in,trim=.25cm .25cm .25cm .15cm,clip]{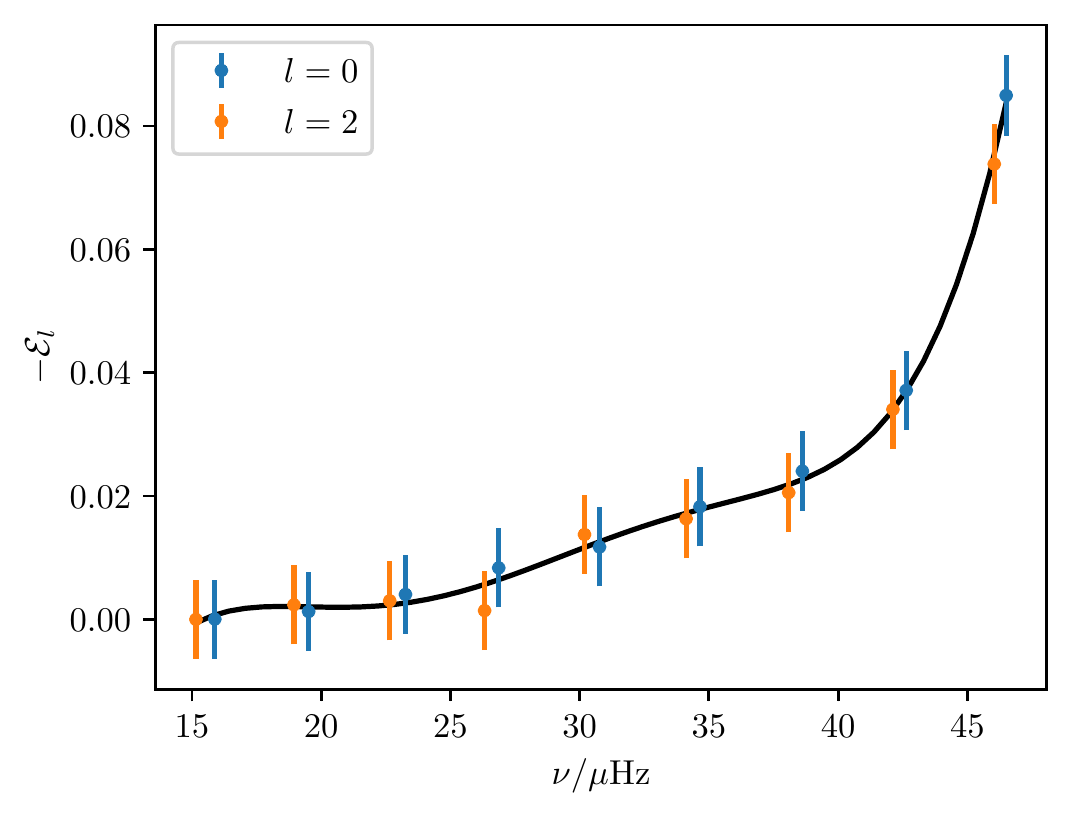}}{
  \node at (.85, .2) {\textbf{(c)}: $\chi^2_\epsilon=0.08$};
  }
  \annotate{\includegraphics[height=2.56in,trim=.25cm .25cm .25cm .15cm,clip]{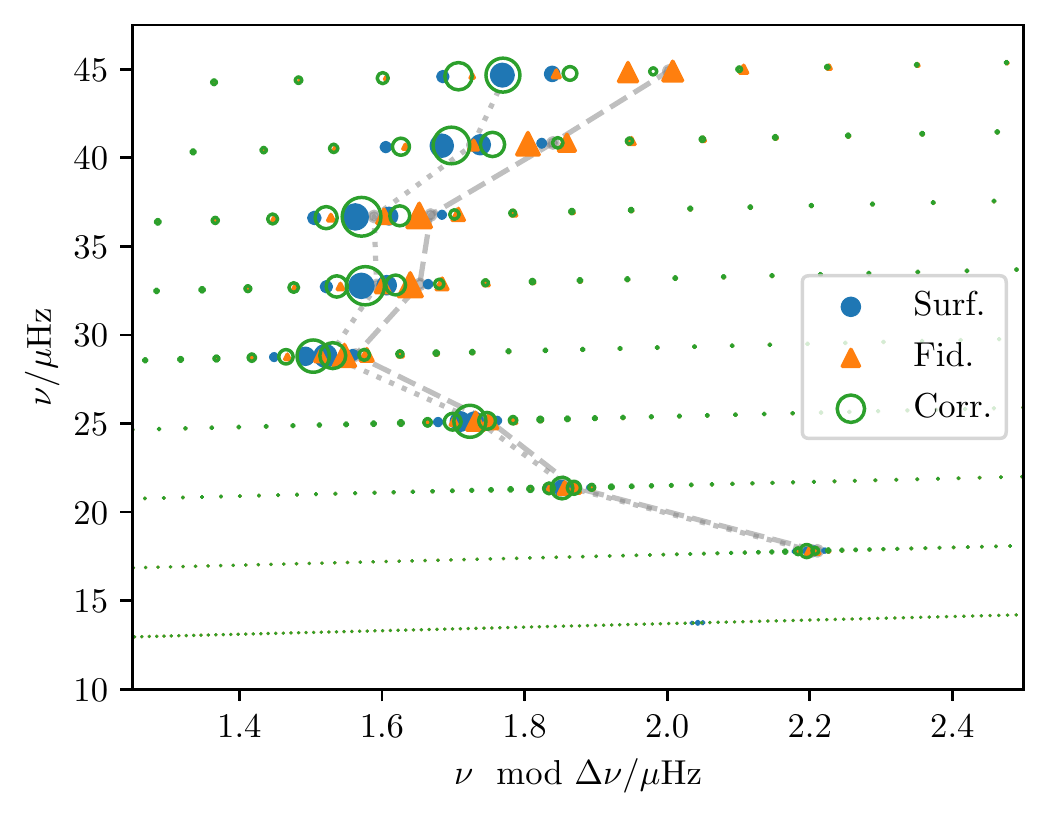}}{
  \node at (.85, .2) {\textbf{(d)}: $\chi^2_\text{tot}=0.54$};
  }
  \caption{Application of different procedures to the perturbed model of \cref{fig:synthetic3}. \textbf{(a)}: $\epsilon$-matching using the most $p$-dominated mixed modes in both the perturbed and fiducial frequency sets. \textbf{(b)}: $\epsilon$-matching of the most $p$-dominated mixed modes of the perturbed frequencies, matched with the fiducial $\pi$ modes. \textbf{(c)}: $\epsilon$-matching procedure restricted to radial and $p$-dominated quadrupole modes, matched with the corresponding model $\pi$ modes, in order to constrain $\mathcal{F}(\nu)$ independently of the dipole modes. \textbf{(d)}: Dipole mixed modes computed via the prescription of \autoref{sec:matrix} after applying the correction implied by $\mathcal{F}(\nu)$ as shown in panel (c).}
  \label{fig:synthetictest3}
\end{figure*}

There are several ways this may be done. For one, we might note that the
most p-dominated mixed mode frequencies from the model are essentially
specified by the \(\pi\) modes. Where the coupling is weak, it is common
practice \citep[as done in e.g.][ for quadrupole mixed
modes]{ball_surface_2018, mckeever_helium_2019, chaplin_age_2020, ong_differential_2021}
to simply match these up with the most p-dominated mixed modes in the
observed set. For the dipole modes, we see from \cref{fig:synthetic3}b
that the g-modes are now sufficiently dense to also permit this
approach. We compare the results of doing this against directly using
the \(\pi\)-modes associated with the model instead (in both cases
matching them up with the most p-dominated of the observed mixed modes)
in \cref{fig:synthetictest3}a and b. Generally speaking, we see that the
use of the model \(\pi\) modes yields clearly superior results, even for
the dipole modes (where the coupling is less weak). This is because the
dipole coupling strength is comparable to the local g-mode spacing.
Consequently it is large enough that even the most p-dominated mixed
mode, i.e.~the one nearest to the corresponding \(\pi\) mode, contains
contributions from multiple g-modes, and hence \(Q > 1\), which inflates
the estimate of \(\mathcal{E}_1\) returned from \cref{eq:modifiedeps}.
This is avoided by using the \(\pi\) modes directly.

On the other hand, restricting ourselves to only the most p-dominated
modes means that we discard information from the nearby, more g-like
mixed modes, which yield better constraints of the interior structure of
the star. In order to include these modes, we once again turn to the
explicit matrix construction. As pointed out in
\citet{ball_surface_2018}, only the very most \(p\)-dominated mixed
modes are observationally accessible at quadrupole or higher degree, and
for these modes the coupling strengths are so weak that the differences
between their frequencies and those of the pure \(\pi\) modes is
trivial, as can be seen in \cref{fig:synthetictest3}c.~Accordingly, in
this regime we may rely on these modes (in addition to the radial modes)
to calibrate \(\mathcal{F}(\nu)\) for use in correcting the dipole
modes, and again feed this into \cref{eq:HEP} via the prescription of
\autoref{sec:matrix}. We show the results of doing this in
\cref{fig:synthetictest3}d.~While the systematic error incurred by this
approximation is still smaller than neglecting mode coupling altogether,
we find that for this evolved red-giant regime the fundamental source of
systematic error is instead the computation of the \(\gamma\)-mode
matrix elements, owing to underlying issues in the usual numerical
scheme (see discussion in \autoref{sec:gammagamma}).

Numerical issues aside, both the evaluation of these matrices, and
solving their associated GHEPs, also become extremely expensive in this
regime. Generally speaking, the coupling strength between any given pair
of \(\pi\) and \(\gamma\) modes increases with the frequency of the
\(\gamma\) mode, since the \(\gamma\) mode eigenfunctions become less
oscillatory with frequency. Accordingly, for such evolved stars, the
matrix elements of a very large number of \(\gamma\) modes has to be
computed, including \(\gamma\) modes of much higher frequency than
typically observed, so that the approximate coupling matrices which we
actually use are sufficiently complete. Moreover, for a pair of
\(N\times N\) matrices, the solution of the corresponding GHEP has a
runtime complexity of \(\mathcal{O}(N^3)\). For this single stellar
model, the computation of these matrix elements took about 4 CPU hours
on an Intel Xeon E5-2670 CPU running at 2.60 GHz. Thus, even if the
numerical issues underlying the computation of the \(\gamma\)-mode
matrix elements are resolved, this procedure appears computationally
impractical to apply at scale for the moment, at least for these evolved
red giants.

\hypertarget{comparison-with-first-order-approximation}{%
\subsubsection{Comparison with first-order
approximation}\label{comparison-with-first-order-approximation}}

In the solar case it is known that, in the absence of mode coupling, the
surface term is well-approximated by a description of the form
\begin{equation}
  \delta\nu_{\text{surf}, nl} \sim f(\nu_{nl}) / I_{nl}.\label{eq:inertiaweighted}
\end{equation} If they account for mode coupling at all, existing
descriptions of the surface term treat mode coupling in precisely this
manner
\citep[e.g.][]{kjeldsen_correcting_2008, ball_correction_2014, perezhernandez_rg_2016}.
In the absence of an a priori constraint on the size of the surface
term, the function \(f(\nu)\) is typically constrained (similarly to
\autoref{sec:matrix}) with reference to only the radial modes, or a
combination of the radial modes and most \(\pi\)-dominated quadrupole
modes. We showed in \autoref{sec:bg14matrix} that using this expression
directly to describe the action of the surface term on the mixed modes,
instead of performing the full matrix computation per
\autoref{sec:matrix}, amounts to truncating the corresponding
perturbative series expansion in powers of \(\lambda\) to first order in
\(\lambda\).

\begin{figure}
\centering
\includegraphics{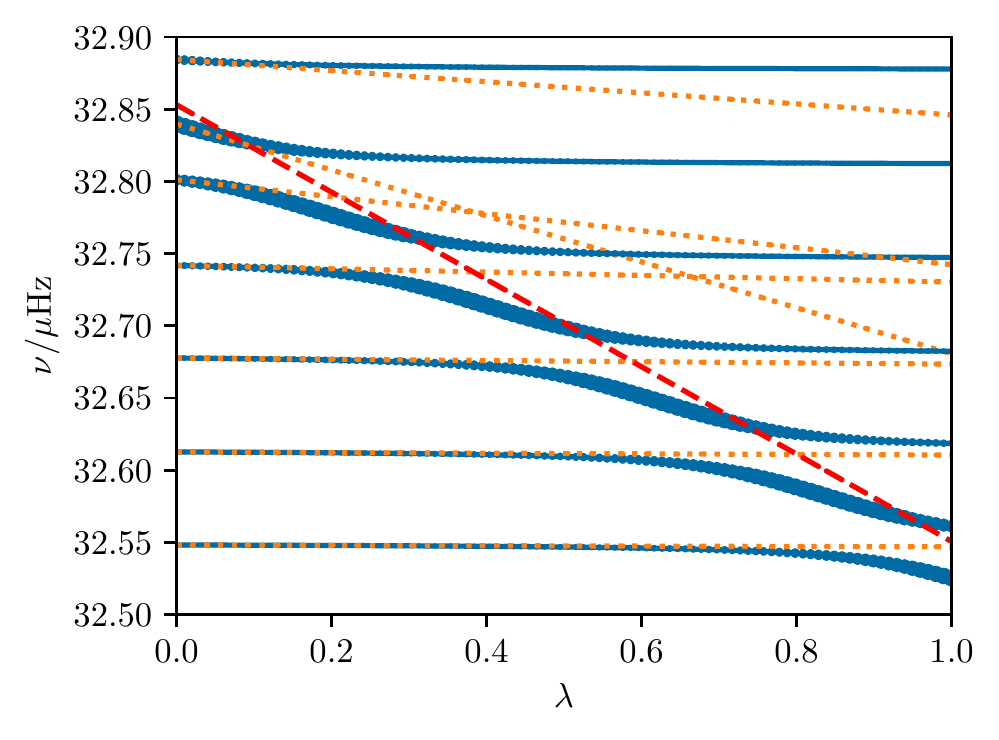}
\caption{Interactions between mode mixing and the surface term, compared
with the first-order approximation to mode coupling. The blue circles
and lines show the frequencies of mixed modes obtained with different
strengths of the surface term (parameterised by \(\lambda\)), while the
orange dotted lines show the predicted mixed-mode frequencies upon
application of an inertia-weighted correction of the form of
\cref{eq:inertiaweighted}. The red dashed line shows the evolution of
the \(\pi\)-mode near these mixed modes, for which
\cref{eq:inertiaweighted} is a much better approximation.
\label{fig:avoided}}
\end{figure}

In \autoref{sec:convergence}, we derived that this perturbative series
fails to converge when the surface term is larger than both the coupling
strength and the local g-mode separation, which is the case for these
evolved red giants. To illustrate the manner in which this approximation
breaks down, we show in \cref{fig:avoided} the effects of applying an
exaggeratedly large synthetic surface perturbation, with parameters
\(A=1, \sigma=0.002\), on the red giant model of the preceding section.

As before, we quantify the strength of the surface term with an
interpolating parameter \(\lambda\) going from 0 to 1. The mixed-mode
frequencies for a corresponding family of interpolated models are shown
with filled blue circles (with sizes given as \(1/Q\)), overlaid on the
solid blue curves. We also show the predicted frequency perturbations
from applying a first-order correction, of the form
\cref{eq:inertiaweighted}, with the orange dotted lines. As expected
from being first-order corrections, these take the form of straight
lines, which are tangent to the blue curves at \(\lambda=0\). These
first-order constructions may potentially change the ordering of the
modes before and after their application \citep[as noted
in][]{ball_surface_2018, li_modelling_2018}. We can see here that this
is a fundamental limitation of first-order corrections in general, since
the corresponding tangent lines may cross freely. Moreover, it is clear
that when the surface term is large enough to change the ordering of the
first-order predictions, these predicted frequencies are entirely
unreliable, even for the most p-dominated modes.

By contrast, the actual mixed modes undergo a series of avoided
crossings, tracing out the trajectory of the underlying \(\pi\)-mode
(shown with the red dashed line), which itself evolves in a linear
fashion as the surface perturbation is ``turned on''. Outside of the
formal radius of convergence (\autoref{sec:convergence}), these avoided
crossings defy linearisation (or, for that matter, approximation by
power series), and are poorly approximated by their tangent lines at
\(\lambda=0\). While the first-order construction still remains
applicable to the bare \(\pi\)-modes, a full account of mode mixing is
in principle mandatory for describing mixed modes in these situations.

At the same time, we have shown in the preceding section that for these
evolved red giants, our existing implementation of the matrix machinery
required for this is at once increasingly susceptible to numerical
error, and prohibitively expensive to actually use. At present, it
appears that we have little choice but to restrict our attention to the
pure \(\pi\)-mode system when dealing with such red giants. While
earlier works have raised concerns about potentially incurring
systematic errors when doing so for dipole modes in particular
\citep{ball_surface_2018, ong_differential_2021}, we have demonstrated
in the preceding section that these are at least smaller than the
typical statistical errors, and far preferable to the naive approach of
using p-dominated mixed modes.

\hypertarget{discussion-and-conclusion}{%
\section{Discussion and Conclusion}\label{discussion-and-conclusion}}

The asteroseismic surface term arises as a consequence of structural
differences between stellar models and actual stars, which in turn
originate from modelling error localised to the near-surface layers. We
have shown that the use of classical variational analysis for deriving
frequency responses to these structural perturbations is implicitly
equivalent to truncating a perturbative series expansion of the
underlying eigenvalue problem to leading order in the perturbation
operator.

This kind of first-order approximation holds good for pure \(p\)-modes,
as well as for the \(\pi\)-mode subsystem of a star exhibiting mixed
modes. However, such an approximation does not describe mixed modes well
when the size of the frequency shift from the surface term is comparable
to the coupling strength between the \(\pi\) and \(\gamma\) mode
cavities. We have derived a more general matrix construction that
remains valid in these cases. We have also derived both a first-order
generalisation and an analogous matrix construction for the
\(\epsilon\)-matching algorithm described in
\citet{roxburgh_asteroseismic_2016}. From both analytic considerations
(\autoref{sec:convergence}) and injection-recovery tests on stellar
models (\autoref{sec:numerics}), we have shown that various sources of
systematic error dominate for stellar models in different evolutionary
stages. In particular we find that:

\begin{itemize}
\tightlist
\item
  Where the surface term is small compared to the coupling strength,
  existing methods may be used with minimal modification, so that mode
  coupling is accounted for to first order;
\item
  Where the surface term is comparable to the coupling strength and/or
  the g-mode separation, such first-order analysis may not be adequate;
\item
  Where the surface term is much larger than both the coupling strength
  and the g-mode separation, first-order constructions yield
  satisfactory results only when restricted to the \(\pi\)-mode
  subsystem. If mixed modes are to be explicitly considered at all, then
  in principle the full matrix computation of \autoref{sec:matrix} must
  be used.
\item
  For very evolved red giant stars, the full matrix construction becomes
  computationally impractical, and in any case we encounter an
  additional source of systematic error, arising from the numerical
  scheme used to compute \(\gamma\)-mode matrix elements at high order.
  In these cases, it appears for the time being that the only practical
  option is to limit attention to the behaviour of the \(\pi\)-mode
  subsystem.
\end{itemize}

One limitation of this analysis is that it is strictly applicable only
to situations where both the fiducial and perturbed wave operator can be
assumed to be Hermitian. By contrast, it is typically assumed that part
of the surface term can be attributed to nonadiabatic wave propagation
near the surface (typically neglected in pulsation calculations), which
results in a non-Hermitian perturbation operator. Our analysis remains
approximately applicable so long as the imaginary components of the
nonadiabatic eigenvalues are small. However, in the more general,
strongly nonadiabatic, case, a full non-Hermitian treatment
\citep[e.g.~generalising the approach of][]{jcd_nonadiabatic_1981} may
ultimately prove necessary.

The inferred values of certain global properties, such as stellar
masses, radii, and ages, are known to be robust to different treatments
of the surface term
\citep{nsamba_asteroseismic_2018, compton_surface_2018, basu_robustness_2018}
on the main sequence, where the observed modes are pure p-modes. They
are, however, sensitive to these methodological choices
\citep{jorgensen_investigating_2020, ong_differential_2021} for evolved
red giants, where the observed modes are close to bare \(\pi\)-modes.
Inferences of other properties, such as their initial helium abundances,
have been shown to be sensitive to the choice of surface term treatment
in both these extreme regimes, although the nature of this sensitivity
exhibits qualitative differences. In Paper II, we examine the
corresponding situation for subgiant stars, which lie in the
intermediate regime.

These considerations are especially relevant given that, owing to
observational limitations, the vast majority of known solar-like
oscillators are evolved stars, and therefore exhibit mode mixing of some
form. On the other hand, essentially all surface-term corrections in the
literature \citep[see e.g.][ for a review]{jorgensen_investigating_2020}
only account for mode mixing to first order, if at all. Given that
almost all stars on the \emph{TESS} short-cadence asteroseismic target
list are subgiants, it is imperative that surface-term systematics of
the kind we studied in \citet{ong_differential_2021} be properly
identified and characterised for subgiants in particular. The
constructions we have presented in this work are essential to this task.

\acknowledgements

We thank U. Banik, B. Mosser, \amend{and the anonymous referee} for
constructive feedback; we also thank R. Townsend for technical
assistance with \gyre. We have made available Python scripts to evaluate
the matrix elements discussed here and in \ob~at
\url{https://gitlab.com/darthoctopus/mesa-tricks}. This work was
partially supported by NASA K2 GO Award 80NSSC19K0102 and NASA TESS GO
Award 80NSSC19K0374.

\software{NumPy \citep{numpy}, SciPy stack \citep{scipy}, AstroPy \citep{astropy:2013,astropy:2018}, Pandas \citep{mckinney-proc-scipy-2010}, \mesa\ \citep{mesa_paper_1,mesa_paper_2,mesa_paper_4}, \gyre\ \citep{townsend_gyre_2013}.}

\appendix

\hypertarget{gamma-mode-frequency-correction}{%
\section{\texorpdfstring{\(\gamma\)-mode frequency
correction}{\textbackslash gamma-mode frequency correction}}\label{gamma-mode-frequency-correction}}

\label{sec:gammagamma} We recall that linear adiabatic oscillations in a
nonrotating star can be expressed as linear combinations of displacement
eigenfunctions \begin{equation}
    \vec{\xi}(r, \theta, \phi, t) = e^{\pm i\omega t}\left(\xi_r(r) \mathbf{Y}_l^m + \xi_h(r)\mathbf{\Psi}_l^m\right)
\end{equation} which emerge as solutions to the system of differential
equations \begin{equation}
    \begin{aligned}
    {1 \over r^2}{\mathrm d \over \mathrm d r} (r^2 \xi_r) - {g \over c_s^2}\xi_r + \left(\alpha_\gamma - {S_l^2 \over \omega^2}\right) {P_1 \over \rho c_s^2} &= {\Lambda^2 \over \omega^2}\Phi_1,\\
    {1 \over \rho} {\mathrm d P_1 \over \mathrm d r} + {g \over c_s^2}P_1 + (\alpha_\pi N^2-\omega^2)\xi_r &= -{\mathrm d \Phi_1 \over \mathrm d r},\\
    {1 \over r^2}{\mathrm d \over \mathrm d r}\left(r^2 {\mathrm d \Phi_1\over \mathrm d r}\right) - {\Lambda^2 \over r^2} \Phi_1 = 4\pi G\rho\left({P_1 \over \rho c_s^2} + {N^2 \over g}\xi_r\right)\\
    \xi_h = {1 \over r \omega^2}\left({P_1\over \rho} + \Phi_1\right).
    \end{aligned}\label{eq:osc}
\end{equation} We adopt the notational convention that all scalar
perturbed quantities admit separation of variables as, e.g.,
\begin{equation}
    \rho'(r, \theta, \phi)=\rho_{lmn}(r) Y_l^m(\theta,\phi).
\end{equation} Under ordinary circumstances,
\(\alpha_\gamma = \alpha_\pi = 1\). In \citet{ong_semianalytic_2020}, we
demonstrated that isolation of the \(\pi\) and \(\gamma\) mode cavities
can be effected by setting \(\alpha_\pi = 0\) or \(\alpha_\gamma= 0\),
respectively. The full set of mixed-mode eigenvalues can then be
recovered by solving the GHEP \cref{eq:HEP}, where the matrix elements
are specified by integrals taken with respect to the \(\pi\) and
\(\gamma\) mode eigenfunctions.

The expression we derived for the coupling matrix elements within the
\(\gamma\gamma\) subspace in our previous paper assumes that the
definition of the tangential component of the displacement wavefunction
is modified. We showed previously that this results in a non-Hermitian
wave operator. This is an undesirable property, and further
investigation reveals it to yield results that are numerically
inconsistent with mixed modes obtained from integrating the full
equations. We now derive a Hermitian expression for this
\(\gamma\gamma\) coupling matrix, following from a different set of
physical assumptions.

Specifically, we express the first line of \cref{eq:osc} in
coordinate-free form by modifying the continuity equation, which we
write as \begin{equation}
    \rho' + \nabla \cdot (\rho \vec{\xi}) = (1 - \alpha_\gamma){P' \over c_s^2},\label{eq:cont}
\end{equation} from which we can eliminate the density via the adiabatic
relation \begin{equation}
    \rho' = {P' \over c_s^2} + \rho \vec{\xi} \cdot \left({1 \over \rho c_s^2} \nabla P - {1 \over \rho} \nabla \rho\right)\label{eq:ad}
\end{equation} to obtain the desired result: \begin{equation}
    \nabla \cdot \vec\xi + \alpha_\gamma {P' \over \rho c_s^2} + {1 \over \rho c_s^2} \vec\xi \cdot \nabla P = 0.\label{eq:cont2}
\end{equation} On the other hand, the time-independent wave operator
acts on the momentum equation: \begin{equation}
    \mathcal{L}\vec{\xi} = -\omega^2 \rho \vec{\xi} = -\nabla P' + \rho' \vec{g} - \rho \nabla \Phi'.\label{eq:waveorig}
\end{equation} As in our last paper, we define the \(\gamma\)-mode wave
operator \(\mathcal L_\gamma\) to be the wave operator whose orthogonal
basis functions are the eigenfunctions of the system of equations
(\cref{eq:osc}) with \(\alpha_\gamma \to 0\); it is related to the full
time-independent wave operator as
\(\mathcal L_\text{tot} \equiv \mathcal L_\gamma + R_\gamma\). Under
ordinary circumstances (\(\alpha_\gamma=1\)), the right-hand-side of
\cref{eq:cont} vanishes, allowing us to eliminate \(\rho'\), \(P'\), and
\(\Phi'\) from \cref{eq:waveorig} in favour of the displacement
eigenfunctions, yielding the usual, manifestly Hermitian expression
\begin{equation}
    \mathcal{L}_\text{tot}\vec{\xi} = \nabla \left(\vec\xi \cdot \nabla P + c_s^2 \rho \nabla \cdot \vec\xi\right) - \vec{g} \nabla \cdot (\rho \vec\xi) - \rho G \nabla\left(\int \mathrm d^3 x' {\nabla \cdot(\rho \vec\xi) \over |x - x'|}\right).\label{eq:wavetot1}
\end{equation} Comparing \cref{eq:wavetot1} with \cref{eq:cont2}, we see
that the first term of \cref{eq:wavetot1} vanishes when the full wave
operator acts on \(\gamma\) modes (so that \(\alpha_\gamma\) = 0).
Consequently, the action of the total wave operator on \(\gamma\) modes
is such that \begin{equation}
    \mathcal{L}_\text{tot}\vec{\xi_\gamma} = -\vec{g} \nabla \cdot (\rho \vec\xi_\gamma) - \rho G \nabla\left(\int \mathrm d^3 x' {\nabla \cdot(\rho \vec\xi_\gamma) \over |x - x'|}\right).\label{eq:wavetot2}
\end{equation} With the displacement wavefunctions normalised so that
\begin{equation}
    \int \mathrm d^3 x \ \rho \vec\xi_{\gamma,i} \cdot \vec\xi_{\gamma,j} = \delta_{ij}
\end{equation} where each of \(i, j\) stands in for the relevant
multi-index \(n,l,m\) of the corresponding mode, we define the matrix
\(\mathbf{L}_{\gamma\gamma}\) with elements specified as
\begin{equation}
\begin{aligned}
    &L_{\gamma\gamma,ij} = \left<\vec\xi_{\gamma,i}, \mathcal{L}\vec\xi_{\gamma,j}\right>
    \\&= \int \mathrm d^3 x\  \vec\xi_{\gamma,i} \cdot \left( -\vec{g} \nabla \cdot (\rho \vec\xi_{\gamma,j}) - \rho G \nabla\left(\int \mathrm d^3 x' {\nabla \cdot(\rho \vec\xi_{\gamma,j}) \over |x - x'|}\right)\right)
    \\&\equiv -\omega_{\gamma,i}^2 \delta_{ij} + R_{\gamma\gamma,ij}. \label{eq:integral1}
\end{aligned}
\end{equation}

On the other hand, we may also perform this kind of elimination directly
on the \(\gamma\)-mode wave operator subject to the modified continuity
equation \cref{eq:cont}. Since the right-hand-side no longer vanishes,
we cannot eliminate the pressure directly. Instead, we obtain
\begin{equation}
    \mathcal L_\gamma \vec \xi_{\gamma} = -\nabla P' + \vec{g} \left({P'\over c_s^2} - \nabla \cdot \left(\rho \xi_\gamma\right)\right) +\rho G \nabla\left(\int \mathrm d^3 x' {\left({P'\over c_s^2} - \nabla \cdot \left(\rho \xi_\gamma\right)\right) \over |x - x'|}\right).
\end{equation} Under the action of the inner product, the first term can
be integrated by parts, yielding a surface integral that vanishes, given
that the displacement wavefunctions for \(\gamma\)-modes also vanish on
the outer boundary. We then obtain \begin{equation}
    \int \mathrm d^3 x\ \vec\xi_{\gamma,i} \cdot \left(-\nabla P'_j + \vec g {P'_j \over c_s^2}\right) = \int \mathrm d^3 x\ P'_j \cdot \left(\nabla \cdot \vec\xi_{\gamma,i} + {1 \over \rho c_s^2}\vec\xi_{\gamma, i} \cdot \nabla P\right) = 0,
\end{equation} by \cref{eq:cont2}. Therefore, within the \(\gamma\)-mode
system, we have an explicit expression for \(R_{\gamma\gamma,ij}\),
which are the matrix elements of
\(\mathcal R_\gamma = \mathcal L_\text{tot} - \mathcal L_\gamma\):
\begin{equation}
    R_{\gamma\gamma,ij} = -G\int \mathrm d^3 x\ \rho \vec\xi_{\gamma,i} \cdot \nabla \left(\int \mathrm d^3 x'\ {P'_j/c_s^2\over |x - x'|}\right).\label{eq:integral2}
\end{equation} We have found these two expressions to be consistent with
each other, and with \cref{eq:Rpi} when used to evaluate
\(R_{\pi\gamma}\).

\hypertarget{systematic-errors}{%
\subsection{Systematic Errors}\label{systematic-errors}}

The primary source of error in this computation depends strongly on the
order of the modes under consideration. It is well-known that numerical
solutions of the boundary-value problem, under the usual boundary
conditions typically used in asteroseismology, yield basis functions
that are only approximately orthonormal when the domain of integration
is bounded. Generally speaking, this is because orthonormality of the
eigenfunctions requires that certain surface integrals vanish when
integrating by parts; however, in practice this is only true under
certain restricted choices of boundary conditions, which are not
typically used for seismic modelling. In deriving the above expressions
we have made similar assumptions, and therefore incur errors of a
similar kind. These systematic errors are small in a relative sense, and
become most significant at low \(n_g\). By contrast, at high \(n_g\),
the integrands become highly oscillatory, and potentially yield
accumulated truncation error which increases with \(n_g\). Again, this
error is small in a relative sense.

\begin{figure}
\centering
\includegraphics{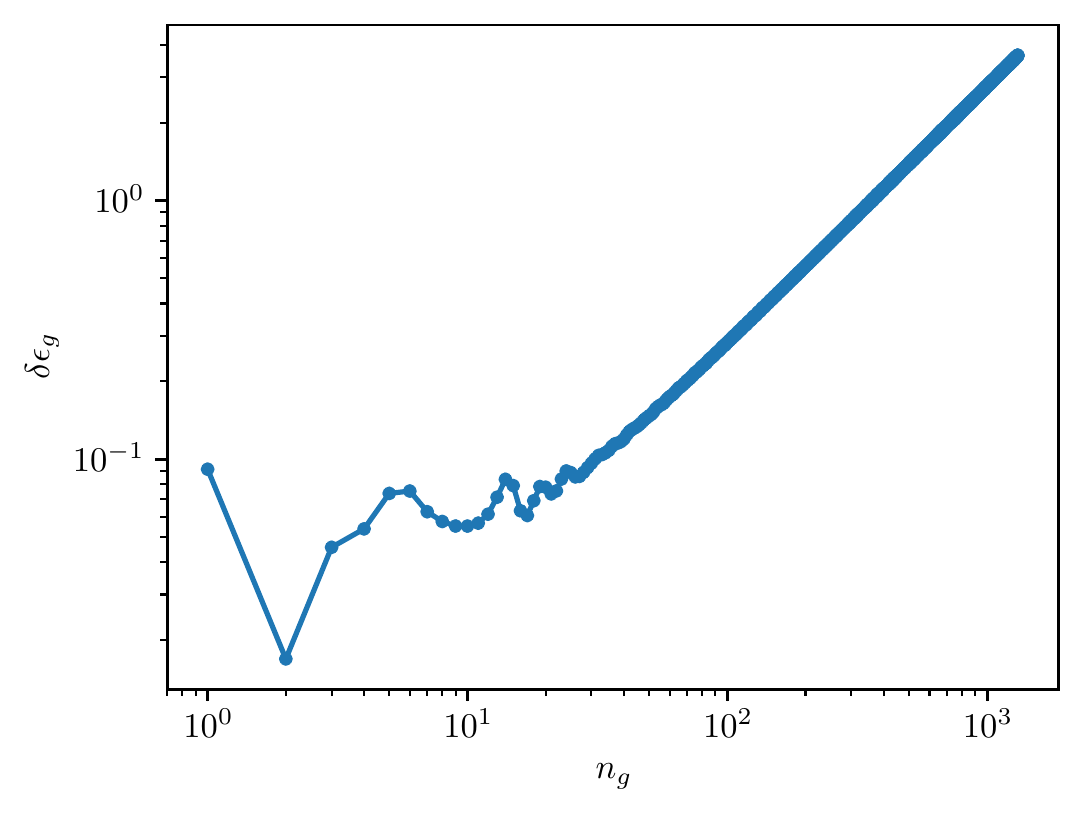}
\caption{Estimate of the truncated absolute integration error in the
g-mode phase function \(\epsilon_g\), as a function of the g-mode radial
order \(n_g\) (see text for complete
description).\label{fig:integrationerror}}
\end{figure}

On the other hand, we require low absolute --- rather than relative ---
integration error, in order to accurately describe the \(\gamma\)-modes.
In particular, given an asymptotic relation of the form \begin{equation}
  P_{l,n_g} \sim \Delta\Pi_l\left(n_g + \epsilon_{g, l}(n_g)\right),\label{eq:error}
\end{equation} we require that \(\epsilon_g\) be estimated accurately in
order to correctly describe the mixed-mode avoided crossings. To
illustrate how both these sources of error scale with \(n_g\), we
estimate the relative error \(\delta P_{n_g} / P_{n_g}\) by evaluating
the inner product \(\left<\vec{\xi}_{n_g}, \vec{\xi}_{n_g + 1}\right>\)
for the \(\gamma\)-modes of the red giant model in
\autoref{evolved-red-giant}; in the absence of both sources of error
this should yield 0 everywhere. This is a relative error estimate
because the mode eigenfunctions are all normalised to yield
\(\left<\vec{\xi}_{n_g}, \vec{\xi}_{n_g}\right> = 1\). We find the
corresponding absolute integration error in \(\epsilon_g\) using
\cref{eq:error}, showing the results in \cref{fig:integrationerror}.

We see that for \(n_g \lesssim 100\), these integration errors are
small. However, as \(n_g\) continues to increase, this absolute error
appears to grow without bound.
\amend{In our numerical implementation, we have fitted a slowly-varying correction to the diagonal elements of the $\gamma$-mode subsystem as a polynomial in the frequency, whose coefficients are chosen to minimise the sum of squared differences between g-dominated mixed-mode frequencies from this matrix construction, compared to those returned directly from GYRE. Even so, \cref{fig:integrationerror} indicates that this strategy is not tenable for $n_g \gtrsim 200$.}

  \bibliography{biblio.bib}

\end{document}

%% file: macros.tex
\usepackage[capitalize]{cleveref}
\usepackage{CJKutf8}

\newcommand{\Dnu}{\ensuremath{\Delta\nu}}
\newcommand{\numax}{\ensuremath{{\nu_\text{max}}}}
\newcommand{\amlt}{\ensuremath{{\alpha_\text{MLT}}}}

\newcommand{\chinesename}{{\begin{CJK}{UTF8}{gbsn}(王加冕)\end{CJK}}}

\newcommand{\amend}[1]{{{#1}}}
\newcommand{\annotate}[2]{\begin{tikzpicture}
    \node[anchor=south west,inner sep=0,align=center] (image) at (0,0) {
    #1
    };
    \begin{scope}[x={(image.south east)},y={(image.north west)}]
    #2
    \end{scope}
\end{tikzpicture}}
\def\bibitem{\vskip\bibskip\footnotesize\savebibitem}

%% file: preamble.tex
\correspondingauthor{Joel Ong}
\email{joel.ong@yale.edu}
\author[0000-0001-7664-648X]{J. M. Joel Ong \chinesename}
\affiliation{Department of Astronomy, Yale University, 52 Hillhouse Ave., New Haven, CT 06511, USA}
\author[0000-0002-6163-3472]{Sarbani Basu}
\affiliation{Department of Astronomy, Yale University, 52 Hillhouse Ave., New Haven, CT 06511, USA}
\author[0000-0002-7403-2764]{Ian W. Roxburgh}
\affiliation{Astronomy Unit, Queen Mary University of London, Mile End Road, London E1 4NS, UK}
\received{June 4, 2021}
\revised{July 5, 2021}
\accepted{July 6, 2021}
\shortauthors{Ong, Basu, Roxburgh}
\def\sectionautorefname{Section}
\def\subsectionautorefname{Section}
\def\subsubsectionautorefname{Section}

